\documentclass{ws-p8-50x6-00}
\usepackage{psfig}
\begin{document}

\makeatletter
\def\ps@plain{\let\@mkboth\@gobbletwo
     \let\@oddhead\@empty\def\@oddfoot{\reset@font\hfil\thepage
     \hfil}\let\@evenhead\@empty\let\@evenfoot\@oddfoot}
\makeatother
\pagestyle{plain}
\def\lesssim{\mathrel{\mathpalette\vereq<}}
\def\gtrsim{\mathrel{\mathpalette\vereq>}}
\makeatletter
\def\vereq#1#2{\lower3pt\vbox{\baselineskip1.5pt \lineskip1.5pt
\ialign{$\m@th#1\hfill##\hfil$\crcr#2\crcr\sim\crcr}}}
\makeatother

\title{Supersymmetry Phenomenology}

\author{Hitoshi Murayama}

\address{Department of Physics, University of California, Berkeley, CA 
94720, USA}

\address{Lawrence Berkeley National Laboratory, Berkeley, CA 94720\\
E-mail: murayama@lbl.gov}  


\maketitle

\abstracts{This is a very pedagogical review of supersymmetry
phenomenology, given at ICTP Summer School in 1999, aimed mostly at
students who had never studied supersymmetry before.  It starts with an
analogy that the reason why supersymmetry is needed is similar to the
reason why the positron exists.  It introduces the construction of
supersymmetric Lagrangians in a practical way.  The low-energy
constraints, renormalization-group analyses, collider
phenomenology, and frameworks of mediating supersymmetry breaking are briefly 
discussed.  }

\section{Motivation}

\subsection{Problems in the Standard Model}

The Standard Model of particle physics, albeit extremely successful
phenomenologically, has been regarded only as a low-energy effective
theory of the yet-more-fundamental theory.  One can list many reasons 
why we think this way, but a few are named below.

\begin{table}
    \centering
        \caption{The fermionic particle content of the Standard Model.  
        Here we've put primes on the neutrinos in the same spirit of 
        putting primes on the down-quarks in the quark doublets, 
        indicating that the mass eigenstates are rotated by the MNS and 
        CKM matrices, respectively.  The subscripts $g,r,b$ refer to 
        colors.}
    \begin{tabular}{llllll}
        $\displaystyle \left( \begin{array}{c} \nu'_{e} \\ e \end{array}
        \right)_{L}^{-1/2}$
        & $\displaystyle \left( \begin{array}{c} \nu'_{\mu} \\ \mu \end{array}
        \right)_{L}^{-1/2}$
        & $\displaystyle \left( \begin{array}{c} \nu'_{\tau} \\ \tau \end{array}
        \right)_{L}^{-1/2}$ &
        $e_{R}^{-1}$ & $\mu_{R}^{-1}$ & $\tau_{R}^{-1}$ \\
        $\displaystyle \left( \begin{array}{c} u \\ d' \end{array}
        \right)_{L,g}^{1/6}$
        & $\displaystyle \left( \begin{array}{c} c \\ s' \end{array}
        \right)_{L,g}^{1/6}$
        & $\displaystyle \left( \begin{array}{c} t \\ b' \end{array}
        \right)_{L,g}^{1/6}$ 
        & $\displaystyle 
        \begin{array}{l} u_{R,g}^{2/3} \\ d_{R,g}^{-1/3} \end{array}$
        & $\displaystyle 
        \begin{array}{l} c_{R,g}^{2/3} \\ s_{R,g}^{-1/3} \end{array}$
        & $\displaystyle 
        \begin{array}{l} t_{R,g}^{2/3} \\ b_{R,g}^{-1/3} \end{array}$ \\
        $\displaystyle \left( \begin{array}{c} u \\ d' \end{array}
        \right)_{L,r}^{1/6}$
        & $\displaystyle \left( \begin{array}{c} c \\ s' \end{array}
        \right)_{L,r}^{1/6}$
        & $\displaystyle \left( \begin{array}{c} t \\ b' \end{array}
        \right)_{L,r}^{1/6}$ 
        & $\displaystyle 
        \begin{array}{l} u_{R,r}^{2/3} \\ d_{R,r}^{-1/3} \end{array}$
        & $\displaystyle 
        \begin{array}{l} c_{R,r}^{2/3} \\ s_{R,r}^{-1/3} \end{array}$
        & $\displaystyle 
        \begin{array}{l} t_{R,r}^{2/3} \\ b_{R,r}^{-1/3} \end{array}$ \\
        $\displaystyle \left( \begin{array}{c} u \\ d' \end{array}
        \right)_{L,b}^{1/6}$
        & $\displaystyle \left( \begin{array}{c} c \\ s' \end{array}
        \right)_{L,b}^{1/6}$
        & $\displaystyle \left( \begin{array}{c} t \\ b' \end{array}
        \right)_{L,b}^{1/6}$ 
        & $\displaystyle 
        \begin{array}{l} u_{R,b}^{2/3} \\ d_{R,b}^{-1/3} \end{array}$
        & $\displaystyle 
        \begin{array}{l} c_{R,b}^{2/3} \\ s_{R,b}^{-1/3} \end{array}$
        & $\displaystyle 
        \begin{array}{l} t_{R,b}^{2/3} \\ b_{R,b}^{-1/3} \end{array}$
        \end{tabular}
    \label{tbl:fermions}
\end{table}
\begin{table}
    \centering
    \caption{The bosonic particle content of the Standard Model.}
    \begin{tabular}{ccc}
        $W^{1}, W^{2}, H^{+}, H^{-}$ & $\longrightarrow$ & $W^{+}, W^{-}$  \\
        $W^{3}, B, {\rm Im} (H^{0})$ & $\longrightarrow$ & $\gamma, Z$  \\
        $g\times 8$ &  &   \\
        ${\rm Re}H^{0}$ & $\longrightarrow$ & $H$
    \end{tabular}
    \label{tbl:bosons}
\end{table}

First of all, the quantum number assignments of the fermions under the 
standard $SU(3)_{C} \times SU(2)_{L} \times U(1)_{Y}$ gauge group 
(Table~\ref{tbl:fermions}) appear utterly bizarre. 
Probably the hypercharges are the weirdest of all.  These assignments, 
however, are crucial to guarantee the cancellation of anomalies which 
could jeopardize the gauge invariance at the quantum level, rendering 
the theory inconsistent.  Another related puzzle is why the 
hypercharges are quantized in the unit of $1/6$.  In principle, the 
hypercharges can be any numbers, even irrational.  However, the 
quantized hypercharges are responsible for neutrality of bulk matter 
$Q(e) + 2 Q(u) + Q(d) = Q(u) + 2 Q(d) = 0$ at a precision of 
$10^{-21}$.\cite{PDG}

The gauge group itself poses a question as well.  Why are there 
seemingly unrelated three independent gauge groups, which somehow 
conspire together to have anomaly-free particle content in a 
non-trivial way?  Why is ``the strong interaction'' strong and ``the weak 
interaction'' weaker?  

The essential ingredient in the Standard Model which appears the
ugliest to most people is the electroweak symmetry breaking.  In the
list of bosons in the Standard Model Table~\ref{tbl:bosons}, the gauge
multiplets are necessary consequences of the gauge theories, and they
appear natural.  They of course all carry spin 1.  However, there is
only one spinless multiplet in the Standard Model: the Higgs doublet
\begin{equation}
    \left( \begin{array}{c} H^{+} \\ H^{0} \end{array} \right)
    \label{eq:SMHiggs}
\end{equation}
which condenses in the vacuum due to the Mexican-hat potential.  It is 
introduced just for the purpose of breaking the electroweak symmetry 
$SU(2)_{L} \times U(1)_{Y} \rightarrow U(1)_{\rm QED}$.  The potential 
has to be arranged in a way to break the symmetry without any 
microscopic explanations.  

Why is there a seemingly unnecessary three-fold repetition of 
``generations''?  Even the second generation led the Nobel Laureate 
I.I. Rabi to ask ``who ordered muon?''  Now we face even more puzzling 
question of having three generations.  And why do the fermions have a 
mass spectrum which stretches over almost six orders of magnitude 
between the electron and the top quark?  This question becomes even 
more serious once we consider the recent evidence for neutrino 
oscillations which suggest the mass of the third-generation 
neutrino $\nu'_{\tau}$ of about $0.05$~eV.\cite{SKatm} 
This makes the mass spectrum stretching over {\it thirteen}\/ orders 
of magnitude.  We have no concrete understanding of the mass spectrum 
nor the mixing patterns.

\subsection{Drive to go to Shorter Distances}

All the puzzles raised in the previous section (and more) cry out for
a more fundamental theory underlying the Standard Model.  What history
suggests is that the fundamental theory lies always at shorter
distances than the distance scale of the problem.  For instance, the
equation of state of the ideal gas was found to be a simple
consequence of the statistical mechanics of free molecules.  The van
der Waals equation, which describes the deviation from the ideal one,
was the consequence of the finite size of molecules and their
interactions.  Mendeleev's periodic table of chemical elements was 
understood in terms of the bound electronic states, Pauli exclusion 
principle and spin.  The existence of varieties of nuclide was due to 
the composite nature of nuclei made of protons and neutrons.  The 
list would go on and on.  Indeed, seeking answers at
more and more fundamental level is the heart of the physical science, 
namely the reductionist approach.  

The distance scale of the Standard Model is given by the size of the 
Higgs boson condensate $v = 250$~GeV.  In natural units, it gives 
the distance scale of $d = \hbar c/v = 0.8 \times 10^{-16}$~cm.  We 
therefore would like to study physics at distance scales shorter than 
this eventually, and try to answer puzzles whose partial list was 
given in the previous section.

Then the idea must be that we imagine the Standard Model to be valid 
down to a distance scale shorter than $d$, and then new physics will 
appear which will take over the Standard Model.  But applying the 
Standard Model to a distance scale shorter than $d$ poses a serious 
theoretical problem.  In order to make this point clear, we first 
describe a related problem in the classical electromagnetism, and 
then discuss the case of the Standard Model later along the same 
line.\cite{INS}

\subsection{Positron Analogue}

In the classical electromagnetism, the only dynamical degrees of 
freedom are electrons, electric fields, and magnetic fields.  When an 
electron is present in the vacuum, there is a Coulomb electric field 
around it, which has the energy of
\begin{equation}
    \Delta E_{\rm Coulomb} = \frac{1}{4\pi \varepsilon_{0}}\frac{e^{2}}{r_{e}}.
    \label{eq:ECoulomb}
\end{equation}
Here, $r_{e}$ is the ``size'' of the electron introduced to cutoff the
divergent Coulomb self-energy.  Since this Coulomb self-energy is
there for every electron, it has to be considered to be a part of the
electron rest energy.  Therefore, the mass of the electron receives an
additional contribution due to the Coulomb self-energy:
\begin{equation}
    (m_{e} c^{2})_{\it obs} = (m_{e}c^{2})_{\it bare} + \Delta E_{\rm 
    Coulomb}.
    \label{eq:self}
\end{equation}
Experimentally, we know that the ``size'' of the electron is small,  
$r_{e} \lesssim 10^{-17}$~cm.  This implies that the self-energy 
$\Delta E$ is greater than 10~GeV or so, and hence the ``bare'' 
electron mass must be negative to obtain the observed mass of the 
electron, with a fine cancellation like
\begin{equation}
    0.511 = -9999.489 + 10000.000 {\rm MeV}.
\end{equation}
Even setting a conceptual problem with a negative mass electron aside, 
such a fine-cancellation between the ``bare'' mass of the electron and 
the Coulomb self-energy appears ridiculous.  In order for such a 
cancellation to be absent, we conclude that the classical 
electromagnetism cannot be applied to distance scales shorter than 
$e^{2}/(4\pi \varepsilon_{0} m_{e} c^{2}) = 2.8\times 10^{-13}$~cm.  
This is a long distance in the present-day particle physics' standard.

The resolution to the problem came from the discovery of the
anti-particle of the electron, the positron, or in other words by
doubling the degrees of freedom in the theory.  The Coulomb
self-energy discussed above can be depicted by a diagram where the
electron emits the Coulomb field (a virtual photon) which is absorbed
later by the electron (the electron ``feels'' its own Coulomb field). 
But now that the positron exists (thanks to Anderson back in 1932),
and we also know that the world is quantum mechanical, one should
think about the fluctuation of the ``vacuum'' where the vacuum
produces a pair of an electron and a positron out of nothing together
with a photon, within the time allowed by the energy-time uncertainty
principle $\Delta t \sim \hbar/\Delta E \sim \hbar/(2 m_{e} c^{2})$. 
This is a new phenomenon which didn't exist in the classical
electrodynamics, and modifies physics below the distance scale $d \sim
c \Delta t \sim \hbar c/(2 m_{e} c^{2}) = 200\times 10^{-13}$~cm. 
Therefore, the classical electrodynamics actually did have a finite
applicability only down to this distance scale, much earlier than $2.8
\times 10^{-13}$~cm as exhibited by the problem of the fine
cancellation above.  Given this vacuum fluctuation process, one should
also consider a process where the electron sitting in the vacuum by
chance annihilates with the positron and the photon in the vacuum
fluctuation, and the electron which used to be a part of the
fluctuation remains instead as a real electron. 
V.~Weisskopf\cite{Weisskopf} calculated this contribution to the
electron self-energy for the first time, and found that it is negative
and cancels the leading piece in the Coulomb self-energy exactly:
\begin{equation}
    \Delta E_{\rm pair} = - \frac{1}{4\pi \varepsilon_{0}}\frac{e^{2}}{r_{e}}.
    \label{eq:Epair}
\end{equation}
After the linearly divergent piece $1/r_{e}$ is canceled, the 
leading contribution in the $r_{e} \rightarrow 0$ limit is given by
\begin{equation}
    \Delta E = \Delta E_{\rm Coulomb} + \Delta E_{\rm pair}
    = \frac{3\alpha}{4\pi} m_{e} c^{2} \log \frac{\hbar}{m_{e} c r_{e}}.
    \label{eq:DeltaE}
\end{equation}
There are two important things to be said about this formula.  First, 
the correction $\Delta E$ is proportional to the electron mass and 
hence the total mass is proportional to the ``bare'' mass of the 
electron,
\begin{equation}
    (m_{e} c^{2})_{\it obs} = (m_{e}c^{2})_{\it bare}
    \left[ 1 + \frac{3\alpha}{4\pi} \log \frac{\hbar}{m_{e} c r_{e} }
    \right].
    \label{eq:self2}
\end{equation}
Therefore, we are talking about the ``percentage'' of the correction, 
rather than a huge additive constant.  Second, the correction depends 
only logarithmically on the ``size'' of the electron.  As a result, 
the correction is only a 9\% increase in the mass even for an electron 
as small as the Planck distance $r_{e} = 1/M_{Pl} = 1.6 \times 
10^{-33}$~cm.  

The fact that the correction is proportional to the ``bare'' mass is 
a consequence of a new symmetry present in the theory with the 
antiparticle (the positron): the chiral symmetry.  In the limit of 
the exact chiral symmetry, the electron is massless and the symmetry 
protects the electron from acquiring a mass from self-energy 
corrections.  The finite mass of the electron breaks the chiral 
symmetry explicitly, and because the self-energy correction should 
vanish in the chiral symmetric limit (zero mass electron), the 
correction is proportional to the electron mass.  Therefore, the 
doubling of the degrees of freedom and the cancellation of the power 
divergences lead to a sensible theory of electron applicable to very 
short distance scales.

\subsection{Supersymmetry}

In the Standard Model, the Higgs potential is given by
\begin{equation}
    V = \mu^{2} |H|^{2} + \lambda |H|^{4},
    \label{eq:V}
\end{equation}
where $v^{2} = \langle H \rangle^{2} = -\mu^{2}/2\lambda = (176~{\rm
GeV})^{2}$.  Because perturbative unitarity requires that $\lambda
\lesssim 1$, $-\mu^{2}$ is of the order of $(100~{\rm GeV})^{2}$. 
However, the mass squared parameter $\mu^{2}$ of the Higgs doublet
receives a quadratically divergent contribution from its self-energy
corrections.  For instance, the process where the Higgs doublets
splits into a pair of top quarks and come back to the Higgs boson
gives the self-energy correction
\begin{equation}
   \Delta \mu^{2}_{\rm top} = - 6 \frac{h_{t}^{2}}{4\pi^{2}} 
   \frac{1}{r_{H}^{2}},
    \label{eq:mu2top}
\end{equation}
where $r_{H}$ is the ``size'' of the Higgs boson, and $h_{t} \approx 1$
is the top quark Yukawa coupling.  Based on the same argument in the
previous section, this makes the Standard Model not applicable below
the distance scale of $10^{-17}$~cm.

The motivation for supersymmetry is to make the Standard Model 
applicable to much shorter distances so that we can hope that answers 
to many of the puzzles in the Standard Model can be given by physics 
at shorter distance scales.\cite{motivation}  In order to do so, 
supersymmetry repeats what history did with the positron: doubling the 
degrees of freedom with an explicitly broken new symmetry.  Then the 
top quark would have a superpartner, stop,\footnote{This is a terrible 
name, which was originally meant to be ``scalar top.''  If 
supersymmetry will be discovered by the next generation collider 
experiments, we should seriously look for better names for the 
superparticles.} whose loop diagram gives another contribution to the 
Higgs boson self energy
\begin{equation}
   \Delta \mu^{2}_{\rm stop} = + 6 \frac{h_{t}^{2}}{4\pi^{2}} 
   \frac{1}{r_{H}^{2}}.
    \label{eq:mu2stop}
\end{equation}
The leading pieces in $1/r_{H}$ cancel between the top and stop 
contributions, and one obtains the correction to be
\begin{equation}
    \Delta \mu^{2}_{\rm top} + \Delta \mu^{2}_{\rm top}
    = -6\frac{h_{t}^{2}}{4\pi^{2}} (m_{\tilde{t}}^{2} - m_{t}^{2})
    \log \frac{1}{r_{H}^{2} m_{\tilde{t}}^{2}}.
    \label{eq:Deltamu2}
\end{equation}

One important difference from the positron case, however, is that the
mass of the stop, $m_{\tilde{t}}$, is unknown.  In order for the
$\Delta \mu^{2}$ to be of the same order of magnitude as the
tree-level value $\mu^{2} = -2\lambda v^{2}$, we need
$m_{\tilde{t}}^{2}$ to be not too far above the electroweak scale. 
Similar arguments apply to masses of other superpartners that couple
directly to the Higgs doublet.  This is the so-called naturalness
constraint on the superparticle masses (for more quantitative 
discussions, see papers\cite{naturalness}).

\subsection{Other Directions}

Of course, supersymmetry is not the only solution discussed in the 
literature to avoid miraculously fine cancellations in the Higgs boson 
mass-squared term.  Technicolor (see a review\cite{technicolor}) 
is a beautiful idea which replaces the Higgs doublet by a composite 
techni-quark condensate.  Then $r_{H} \sim 1$~TeV is a truly physical 
size of the Higgs doublet and there is no need for fine cancellations.  
Despite the beauty of the idea, this direction has had problems with 
generating fermion masses, especially the top quark mass, in a way 
consistent with the constraints from the flavor-changing neutral 
currents.  The difficulties in the model building, however, do not 
necessarily mean that the idea itself is wrong; indeed still efforts 
are being devoted to construct realistic models.

Another recent idea is to lower the Planck scale down to the TeV scale 
by employing large extra spatial dimensions.\cite{ADD}  This is a new 
direction which has just started, and there is an intensive activity to 
find constraints on the idea as well as on model building.  Since the 
field is still new, there is no ``standard'' framework one can discuss 
at this point, but this is no surprise given the fact that 
supersymmetry is still evolving even after almost two decades of 
intense research.

One important remark about all these ideas is that they inevitably 
predict interesting signals at TeV-scale collider experiments.  While 
we only discuss supersymmetry in this lecture, it is likely that nature 
has a surprise ready for us; maybe none of the ideas discussed so far 
is right.  Still we know that there is {\it something}\/ out there to 
be uncovered at TeV scale energies.

\section{Supersymmetric Lagrangian}

We do not go into full-fledged formalism of supersymmetric Lagrangians 
in this lecture but rather confine ourselves to a practical introduction 
of how to write down Lagrangians with explicitly broken supersymmetry 
which still fulfill the motivation for supersymmetry discussed in the 
previous section.  One can find useful discussions as well as an 
extensive list of references in a nice review by Steve Martin.\cite{Martin}

\subsection{Supermultiplets}

Supersymmetry is a symmetry between bosons and fermions, and hence 
necessarily relates particles with different spins.  All particles in 
supersymmetric theories fall into supermultiplets, which have both 
bosonic and fermionic components.  There are two types of 
supermultiplets which appear in renormalizable field theories: chiral 
and vector supermultiplets.

Chiral supermultiplets are often denoted by the symbol $\phi$, which can be 
(for the purpose of this lecture) regarded as a short-handed notation 
for the three fields: a complex scalar field $A$, a Weyl fermion 
$\frac{1-\gamma_{5}}{2}\psi = \psi$, and a non-dynamical (auxiliary) 
complex field $F$.  Lagrangians for chiral supermultiplets consist of two 
parts, K\"ahler potential and superpotential.  The K\"ahler potential 
is nothing but the kinetic terms for the fields, usually written with 
a short-hand notation $\int d^{4} \theta \phi^{*} \phi$, which can be 
explicitly written down as
\begin{equation}
    {\cal L} \supset
    \int d^{4} \theta \phi_{i}^{*} \phi_{i} = \partial_{\mu} A^{*}_{i} 
    \partial^{\mu} A_{i} + \bar{\psi}_{i} i\gamma^{\mu} \partial_{\mu} \psi_{i}
    + F^{*}_{i} F_{i}.
    \label{eq:Kahler}
\end{equation}
Note that the field $F$ does not have derivatives in the Lagrangian 
and hence is not a propagating field.  One can solve for $F_{i}$ 
explicitly and eliminate it from the Lagrangian completely.  

The superpotential is defined by a holomorphic function $W(\phi)$ of 
the chiral supermultiplets $\phi_{i}$.  A short-hand notation $\int d^{2} 
\theta W(\phi)$ gives the following terms in the Lagrangian,
\begin{equation}
    {\cal L} \supset
    - \int d^{2} \theta W(\phi) 
    = - \frac{1}{2} \left.\frac{\partial^{2}W}{\partial \phi_{i} \partial 
    \phi_{j}}\right|_{\phi_{i} = A_{i}} \psi^{i} \psi^{j} 
    + \left.\frac{\partial W}{\partial \phi_{i}}\right|_{\phi_{i} = A_{i}} F_{i}.
    \label{eq:superpotential}
\end{equation}
The first term describes Yukawa couplings between fermionic and 
bosonic components of the chiral supermultiplets.  Using both 
Eqs.~(\ref{eq:Kahler}) and (\ref{eq:superpotential}), we can solve for 
$F$ and find
\begin{equation}
    F_{i}^{*} = 
    - \left.\frac{\partial W}{\partial \phi_{i}}\right|_{\phi_{i} = A_{i}}.
    \label{eq:F}
\end{equation}
Substituting it back to the Lagrangian, we eliminate $F$ and instead 
find a potential term
\begin{equation}
    {\cal L} \supset
    - V_{F} = - \left| \frac{\partial W}{\partial \phi_{i}} 
    \right|_{\phi_{i} = A_{i}}^{2}.
    \label{eq:VF}
\end{equation}

Vector supermultiplets $W_{\alpha}$ ($\alpha$ is a spinor index, but never 
mind), which are supersymmetric generalization of the gauge fields, 
consist also of three components, a Weyl fermion (gaugino) $\lambda$, 
a vector (gauge) field $A_{\mu}$, and a non-dynamical (auxiliary) real 
scalar field $D$, all in the adjoint representation of the gauge group 
with the index $a$.  A short-hand notation of their kinetic terms is
\begin{equation}
    {\cal L} \supset
    \int d^{2}\theta W_{\alpha}^{a} W^{\alpha a}
    = -\frac{1}{4} F_{\mu\nu} + \bar{\lambda}^{a} i {\not \!\! D} 
    \lambda^{a} + \frac{1}{2} D^{a} D^{a}.
    \label{eq:WW}
\end{equation}
Note that the field $D$ does not have derivatives in the Lagrangian 
and hence is not a propagating field.  One can solve for $D^{a}$ 
explicitly and eliminate it from the Lagrangian completely.  

Since the vector supermultiplets contain gauge fields, chiral 
supermultiplets which transform non-trivially under the gauge group 
should also couple to the vector multiplets to make the Lagrangian 
gauge invariant.  This requires the modification of the K\"ahler 
potential $\int d^{4} \theta \phi^{*} \phi$ to $\int d^{4} \theta 
\phi^{\dagger} e^{2gV} \phi$, where $V$ is another short-hand notation 
of the vector multiplet.  Then the kinetic terms in 
Eq.~(\ref{eq:Kahler}) are then modified to
\begin{eqnarray}
    & &{\cal L} \supset
    \int d^{4} \theta \phi_{i}^{\dagger} e^{2gV} \phi_{i} \nonumber \\
    && = D_{\mu} A^{\dagger}_{i} D^{\mu} A_{i} 
    + \bar{\psi}_{i} i\gamma^{\mu} D_{\mu} \psi_{i}
    + F^{\dagger}_{i} F_{i} 
    - \sqrt{2} g (A^{\dagger} T^{a} \lambda^{a} \psi)
    - g A^{\dagger} T^{a} D^{a} A. \nonumber \\
    \label{eq:Kahler2}
\end{eqnarray}
Using Eqs.~(\ref{eq:WW},\ref{eq:Kahler2}), one can solve for $D^{a}$ 
and eliminate it from the Lagrangian, finding a potential term
\begin{equation}
    {\cal L} \supset - V_{D} = - \frac{g^{2}}{2} (A^{\dagger} T^{a} A)^{2}
    \label{eq:VD}
\end{equation}
General supersymmetric Lagrangians are given by 
Eqs.~(\ref{eq:Kahler2},\ref{eq:VF},\ref{eq:VD}).\footnote{We dropped 
one possible term called Fayet--Illiopoulos $D$-term possible for 
vector supermultiplets of Abelian gauge groups.  They are often not 
useful in phenomenological models, but there are 
exceptions.\cite{DY,LMSSM}}

Even though we do not go into formal discussions of supersymmetric 
field theories, one important theorem must be quoted: the 
non-renormalization theorem of the superpotential.  Under the 
renormalization of the theories, the superpotential does not receive 
renormalization at all orders in perturbation theory.\footnote{There are 
non-perturbative corrections to the superpotential, however.  See, 
{\it e.g.}\/, a review.\cite{IS}}  We will come back to the virtues of this 
theorem later on.

Finally, let us study a very simple example of superpotential to gain 
some intuition.  Consider two chiral supermultiplets $\phi_{1}$ and 
$\phi_{2}$, with a superpotential
\begin{equation}
    W = m \phi_{1} \phi_{2}.
    \label{eq:Wmass}
\end{equation}
Following the above prescription, the fermionic components have the 
Lagrangian
\begin{equation}
    {\cal L} \supset
    - \frac{1}{2} \frac{\partial^{2}W}{\partial \phi_{i} \partial 
    \phi_{j}} \psi^{i} \psi^{j} = - m \psi_{1} \psi_{2},
    \label{eq:psimass}
\end{equation}
while the scalar potential term Eq.~(\ref{eq:VF}) gives
\begin{equation}
    {\cal L} \supset
    - \left| \frac{\partial W}{\partial \phi_{i}} \right|_{\phi_{i} = A_{i}}^{2}
    = - m^{2} |A_{1}|^{2} - m^{2} |A_{2}|^{2}.
    \label{eq:Amass}
\end{equation}
Obviously, the terms Eqs.~(\ref{eq:psimass},\ref{eq:Amass}) are mass 
terms for the fermionic (Dirac fermion) and scalar components (two 
complex scalars) of the chiral supermultiplets, with the same mass 
$m$.  In general, fermionic and bosonic components in the same 
supermultiplets are degenerate in supersymmetric theories.

\section{Softly Broken Supersymmetry}

We've discussed supersymmetric Lagrangians in the previous section, 
which always give degenerate bosons and fermions.  In the real world, 
we do not see such degenerate particles with the opposite statistics.  
Therefore supersymmetry must be broken.  We will come back later to 
briefly discuss various mechanisms which break supersymmetry 
spontaneously in manifestly supersymmetric theories.  In the 
low-energy effective theories, however, we can just add terms to 
supersymmetric Lagrangians which break supersymmetry explicitly.  The 
important constraint is that such explicit breaking terms should not 
spoil the motivation discussed earlier, namely to keep the Higgs 
mass-squared only logarithmically divergent.  Such explicit breaking 
terms of supersymmetry are called ``soft'' breakings.

The possible soft breaking terms have been classified.\cite{GG}  In 
a theory with a renormalizable superpotential
\begin{equation}
    W = \frac{1}{2} \mu_{ij} \phi_{i} \phi_{j} 
    + \frac{1}{6} \lambda_{ijk} \phi_{i} \phi_{j} \phi_{k},
    \label{eq:Wrenormalizable}
\end{equation}
the possible soft supersymmetry breaking terms have the following forms:
\begin{equation}
    m^{2}_{ij} A_{i}^{*} A_{j}, \qquad
    M \lambda \lambda, \qquad
    \frac{1}{2} b_{ij} \mu_{ij} A_{i} A_{j}, \qquad
    \frac{1}{6} a_{ijk} \lambda_{ijk} A_{i} A_{j} A_{k}.
    \label{eq:soft}
\end{equation}
The first one is the masses for scalar components in the chiral 
supermultiplets, which remove degeneracy between the scalar and spinor 
components.  The next one is the masses for gauginos which remove 
degeneracy between gauginos and gauge bosons.  Finally the last two 
ones are usually called bilinear and trilinear soft breaking terms 
with parameters $b_{ij}$ and $a_{ijk}$ with mass dimension one.  

In principle, any terms with couplings with positive mass dimensions 
are candidates of soft supersymmetry breaking terms.\cite{general}  
Possibilities in theories without gauge singlets are
\begin{equation}
    \psi_{i} \psi_{j}, \qquad 
    A_{i}^{*} A_{j} A_{k}, \qquad
    \psi_{i} \lambda^{a}
    \label{eq:general}
\end{equation}
Obviously, the first term is possible only in theories with 
multiplets with vector-like gauge quantum numbers, and the last term 
with chiral supermultiplets in the adjoint representation.  In the 
presence of gauge singlet chiral supermultiplets, however, such terms 
cause power divergences and instabilities, and hence are not soft in 
general.  On the other hand, the Minimal Supersymmetric Standard 
Model, for instance, does not contain any gauge singlet chiral 
supermultiplets and hence does admit first two possible terms in 
Eq.~(\ref{eq:general}).  There has been some revived interest in 
these general soft terms.\cite{nonstandard}  We will not consider 
these additional terms in the rest of the discussions.  It is also 
useful to know that terms in Eq.~(\ref{eq:soft}) can also induce 
power divergences in the presence of light gauge singlets and 
heavy multiplets.\cite{PS}

It is instructive to carry out some explicit calculations of Higgs 
boson self-energy in supersymmetric theories with explicit soft 
supersymmetry breaking terms.  Let us consider the coupling of the 
Higgs doublet chiral supermultiplet $H$ to left-handed $Q$ and 
right-handed $T$ chiral supermultiplets,\footnote{As will be explained 
in the next section, the right-handed spinors all need to be 
charged-conjugated to the left-handed ones in order to be part of the 
chiral supermultiplets.  Therefore the chiral supermultiplet $T$ 
actually contains the left-handed Weyl spinor $(t_{R})^{c}$.  The 
Higgs multiplet here will be denoted $H_{u}$ in later sections.}  
given by the superpotential term
\begin{equation}
    W = h_{t} Q T H_{u}.
    \label{eq:ht}
\end{equation}
This superpotential term gives rise to terms in the 
Lagrangian\footnote{We dropped terms which do not contribute to the 
Higgs boson self-energy at the one-loop level.}
\begin{equation}
    {\cal L} \supset -h_{t} Q T H_{u}
    - h_{t}^{2} |\tilde{Q}|^{2}|H_{u}|^{2} - h_{t}^{2}|\tilde{T}|^{2} |H_{u}|^{2}
    - m^{2}_{Q} |\tilde{Q}|^{2} - m^{2}_{T} |\tilde{T}|^{2}
    - h_{t} A_{t} \tilde{Q} \tilde{T} H_{u},
    \label{eq:htL}
\end{equation}
where $m^{2}_{Q}$, $m^{2}_{T}$, and $A_{t}$ are soft parameters.  Note 
that the fields $Q$, $T$ are spinor and $\tilde{Q}$, $\tilde{T}$, 
$H_{u}$ are scalar components of the chiral supermultiplets (an 
unfortunate but common notation in the literature).  This explicit 
Lagrangian allows us to easily work out the one-loop self-energy 
diagrams for the Higgs doublet $H_{u}$, after shifting the field 
$H_{u}$ around its vacuum expectation value (this also generates mass 
terms for the top quark and the scalars which have to be consistently 
included).  The diagram with top quark loop from the first term in 
Eq.~(\ref{eq:htL}) is quadratically divergent (negative).  The 
contractions of $\tilde{Q}$ or $\tilde{T}$ in the next two terms also 
generate (positive) contributions to the Higgs self-energy.  In the 
absence of soft parameters $m_{Q}^{2}=m_{T}^{2}=0$, these two 
contributions precisely cancel with each other, consistent with the 
non-renormalization theorem which states that no mass terms 
(superpotential terms) can be generated by renormalizations.  However, 
the explicit breaking terms $m^{2}_{Q}$, $m^{2}_{T}$ make the 
cancellation inexact.  With a simplifying assumption 
$m_{Q}^{2}=m_{T}^{2}=\tilde{m}^{2}$, we find
\begin{equation}
    \delta m_{H}^{2} = - \frac{6 h_{t}^{2}}{(4\pi)^{2}} \tilde{m}^{2}
    \log \frac{\Lambda^{2}}{\tilde{m}^{2}}.
    \label{eq:deltamH}
\end{equation}
Here, $\Lambda$ is the ultraviolet cutoff of the one-loop diagrams.  
Therefore, these mass-squared parameters are indeed ``soft'' in the 
sense that they do not produce power divergences.  Similarly, the 
diagrams with two $h_{t} A_{t}$ couplings with scalar top loop produce 
only a logarithmic divergent contribution.

\section{The Minimal Supersymmetric Standard Model}

Encouraged by the discussion in the previous section that the 
supersymmetry can be explicitly broken while retaining the absence of 
power divergences, we now try to promote the Standard Model to a 
supersymmetric theory.  The Minimal Supersymmetric Standard Model 
(MSSM) is a supersymmetric version of the Standard Model with the 
minimal particle content.

\subsection{Particle Content}

The first task is to promote all fields in the Standard Model to 
appropriate supermultiplets.  This is obvious for the gauge bosons: 
they all become vector multiplets.  For the quarks and leptons, we 
normally have left-handed and right-handed fields in the Standard 
Model.  In order to promote them to chiral supermultiplets, however, 
we need to make all fields left-handed Weyl spinors.  This can be 
done by charge-conjugating all right-handed fields.  Therefore, when we 
refer to supermultiplets of the right-handed down quark, say, we are 
actually talking about chiral supermultiplets whose left-handed 
spinor component is the left-handed anti-down quark field.  As for 
the Higgs boson, the field Eq.~(\ref{eq:SMHiggs}) in the Standard 
Model can be embedded into a chiral supermultiplet $H_{u}$.  It can 
couple to the up-type quarks and generate their masses upon the 
symmetry breaking.  In order to generate down-type quark masses, 
however, we normally use
\begin{equation}
    i\sigma_{2} H^{*} = 
    \left( \begin{array}{c} H^{+} \\ H^{0} \end{array} \right)
    = \left( \begin{array}{c} H^{0*} \\ - H^{-} \end{array} \right).
    \label{eq:Higgs*}
\end{equation}
Unfortunately, this trick does not work in a supersymmetric fashion 
because the superpotential $W$ must be a holomorphic function of the 
chiral supermultiplets and one is not allowed to take a complex 
conjugation of this sort.  Therefore, we need to introduce another 
chiral supermultiplet $H_{d}$ which has the same gauge quantum 
numbers of $i\sigma_{2} H^{*}$ above.\footnote{Another reason to need 
both $H_{u}$ and $H_{d}$ chiral supermultiplets is to cancel the 
gauge anomalies arising from their spinor components.}

In all, the chiral supermultiplets in the Minimal Supersymmetric 
Standard Model are listed in Table~\ref{tbl:MSSM}.

\begin{table}
    \centering
    \caption{The chiral supermultiplets in the Minimal Supersymmetric 
    Standard Model..  The numbers in the bold face refer to 
    $SU(3)_{C}$, $SU(2)_{L}$ representations.  The superscripts are 
    hypercharges.}
    \begin{tabular}{lll}
        $L_{1}({\bf 1},{\bf 2})^{-1/2}$ & $L_{2}({\bf 1},{\bf 2})^{-1/2}$ 
        & $L_{3}({\bf 1},{\bf 2})^{-1/2}$\\
        $E_{1}({\bf 1},{\bf 1})^{+1}$ & $E_{2}({\bf 1},{\bf 1})^{+1}$ 
        & $E_{3}({\bf 1},{\bf 1})^{+1}$ \\
        \hline
        $Q_{1}({\bf 3},{\bf 2})^{1/6}$ & $Q_{2}({\bf 3},{\bf 2})^{1/6}$ 
        & $Q_{3}({\bf 3},{\bf 2})^{1/6}$\\
        $U_{1}({\bf 3},{\bf 1})^{-2/3}$ & $U_{2}({\bf 3},{\bf 1})^{-2/3}$ 
        & $U_{3}({\bf 3},{\bf 1})^{-2/3}$\\
        $D_{1}({\bf 3},{\bf 1})^{+1/3}$ & $D_{2}({\bf 3},{\bf 1})^{+1/3}$ 
        & $D_{3}({\bf 3},{\bf 1})^{+1/3}$\\
        \hline
        & $H_{u}({\bf 1},{\bf 2})^{+1/2}$ & \\
        & $H_{d}({\bf 1},{\bf 2})^{-1/2}$ &
        \end{tabular}
    \label{tbl:MSSM}
\end{table}

The particles in the MSSM are referred to as follows.\footnote{When I
first learned supersymmetry, I didn't believe it at all.  Doubling the
degrees of freedom looked too much to me, until I came up with my own
argument at the beginning of the lecture.  The funny names for the
particles were yet another reason not to believe in it.  It doesn't
sound scientific.  Once supersymmetry will be discovered, we
definitely need better sounding names!} First of all, all quarks,
leptons are called just in the same way as in the Standard Model,
namely electron, electron-neutrino, muon, muon-neutrino, tau,
tau-neutrino, up, down, strange, charm, bottom, top.  Their
superpartners, which have spin 0, are named with ``s'' at the
beginning, which stand for ``scalar.''  They are denoted by the same
symbols as their fermionic counterpart with the tilde.  Therefore, the
superpartner of the electron is called ``selectron,'' and is written
as $\tilde{e}$.  All these names are funny, but probably the worst one
of all is the ``sstrange'' ($\tilde{s}$), which I cannot pronounce at
all.  Superpartners of quarks are ``squarks,'' and those of leptons
are ``sleptons.''  Sometimes all of them are called together as
``sfermions,'' which does not make sense at all because they are bosons. 
The Higgs doublets are denoted by capital $H$, but as we will see
later, their physical degrees of freedom are $h^{0}$, $H^{0}$, $A^{0}$
and $H^{\pm}$.  Their superpartners are called ``higgsinos,'' written
as $\tilde{H}_{u}^{0}$, $\tilde{H}_{u}^{+}$, $\tilde{H}_{d}^{-}$,
$\tilde{H}_{d}^{0}$.  In general, fermionic superpartners of boson in
the Standard Model have ``ino'' at the end of the name.  Spin 1/2
superpartners of the gauge bosons are ``gauginos'' as mentioned in the
previous section, and for each gauge groups: gluino for gluon, wino
for $W$, bino for $U(1)_{Y}$ gauge boson $B$.  As a result of the
electroweak symmetry breaking, all neutral ``inos'', namely two
neutral higgsinos, the neutral wino $\tilde{W}^{3}$ and the bino
$\tilde{B}$ mix with each other to form four Majorana fermions.  They
are called ``neutralinos'' $\tilde{\chi}_{i}^{0}$ for $i=1,2,3,4$. 
Similarly, the charged higgsinos $\tilde{H}_{u}^{+}$,
$\tilde{H}_{d}^{-}$, $\tilde{W}^{-}$, $\tilde{W}^{+}$ mix and form two
massive Dirac fermions ``charginos'' $\tilde{\chi}_{i}^{\pm}$ for
$i=1,2$.  All particles with tilde do not exist in the
non-supersymmetric Standard Model.  Once we introduce $R$-parity in a
later section, the particles with tilde have odd $R$-parity.

\subsection{Superpotential}

The $SU(3)_{C} \times SU(2)_{L} \times U(1)_{Y}$ gauge invariance 
allows the following terms in the superpotential
\begin{eqnarray}
    W & = & \lambda_{u}^{ij} Q_{i} U_{j} H_{u} 
    + \lambda_{d}^{ij} Q_{i} D_{j} H_{d} 
    + \lambda_{e}^{ij} L_{i} E_{j} H_{d} + \mu H_{u} H_{d}
    \nonumber  \\
    &  & + \lambda_{u}^{\prime ijk} U_{i} D_{j} D_{k}
    + \lambda_{d}^{\prime ijk} Q_{i} D_{j} L_{k}
    + \lambda_{e}^{\prime ijk} L_{i} E_{j} L_{k}
    + \mu'_{i} L_{i} H_{u}.
    \label{eq:WMSSM}
\end{eqnarray}
The first three terms correspond to the Yukawa couplings in the 
Standard Model (with exactly the same number of parameters).  The 
subscripts $i,j,k$ are generation indices.  The parameter $\mu$ has 
mass dimension one and gives a supersymmetric mass to both fermionic 
and bosonic components of the chiral supermultiplets $H_{u}$ and 
$H_{d}$.  The terms in the second line of Eq.~(\ref{eq:WMSSM}) are in 
general problematic as they break the baryon ($B$) or lepton ($L$) 
numbers.

If the superpotential contains both $B$- and $L$-violating terms, such 
as $\lambda_{u}^{\prime 112}U_{1}D_{1}D_{2}$ and $\lambda_{d}^{\prime 
121}Q_{1} D_{2} L_{1}$, one can exchange $\tilde{D}_{2} = \tilde{s}$ 
to generate a four-fermion operator
\begin{equation}
        \frac{\lambda_{u}^{\prime 112} \lambda_{d}^{\prime 121}}{m^{2}_{\tilde{s}}}
        (u_{R} d_{R}) (Q_{1} L_{1}),
        \label{eq:dangerous}
\end{equation}
where the spinor indices are contracted in each parentheses and the 
color indices by the epsilon tensor.  Such an operator would 
contribute to the proton decay process $p \rightarrow e^{+} \pi^{0}$ 
at a rate of $\Gamma \sim \lambda^{\prime 4} 
m_{p}^{5}/m_{\tilde{s}}^{4}$, and hence the partial lifetime of the 
order of
\begin{equation}
        \tau_{p} \sim 6 \times 10^{-13}~{\rm sec} \left( 
        \frac{m_{\tilde{s}}}{\rm 1~TeV}\right)^{4} \frac{1}{\lambda^{\prime 4}}.
        \label{eq:taup}
\end{equation}
Recall that the experimental limit on the proton partial lifetime in 
this mode is $\tau_{p} > 1.6 \times 10^{33}$~years.\cite{e+pi0}
Unless the coupling constants are extremely small, this is clearly a 
disaster.

\subsection{$R$-parity}

To avoid this problem of too-rapid proton decay, a common assumption 
is a discrete symmetry called $R$-parity\cite{Rparity} (or matter 
parity\cite{matterparity}).  The $Z_{2}$ discrete charge is given by
\begin{equation}
        R_{p} = (-1)^{2s+3B+L}
        \label{eq:Rp}
\end{equation}
where $s$ is the spin of the particle.  Under $R_{p}$, all standard 
model particles, namely quarks, leptons, gauge bosons, and Higgs 
bosons, carry even parity, while their superpartners odd due to 
the $(-1)^{2s}$ factor.  Once this discrete symmetry is imposed, all 
terms in the second line of Eq.~(\ref{eq:WMSSM}) will be forbidden, 
and we do not generate a dangerous operator such as that in 
Eq.~(\ref{eq:dangerous}).  Indeed, $B$- and $L$-numbers are now 
accidental symmetries of the MSSM Lagrangian as a consequence of the 
supersymmetry, gauge invariance, renormalizability and $R$-parity 
conservation.  

One immediate consequence of the conserved $R$-parity is that the 
lightest particle with odd $R$-parity, {\it i.e.}\/, the Lightest 
Supersymmetric Particle (LSP), is stable.  Another consequence is that 
one can produce (or annihilate) superparticles only pairwise.  These 
two points have important implications on the collider phenomenology 
and cosmology.  Since the LSP is stable, its cosmological relic is a 
good (and arguably the best) candidate for the Cold Dark Matter 
particles (see, {\it e.g.}\/, a review\cite{LSPCDM} on this subject).  
If so, we do not want it to be electrically charged and/or strongly 
interacting; otherwise we should have detected them already.  Then the 
LSP should be a superpartner of $Z$, $\gamma$, or neutral Higgs bosons 
or their linear combination (called neutralino).\footnote{A sneutrino 
can in principle be the LSP,\cite{LMSSM}, but it cannot be the CDM to 
avoid constraints from the direct detection experiment for the CDM 
particles.\cite{snuCDM} It becomes a viable candidate again if there 
is a large lepton number violation.\cite{snu}} On the other hand, the 
superparticles can be produced only in pairs and they decay eventually 
into the LSP, which escapes detection.  This is why the typical 
signature of supersymmetry at collider experiments is the missing 
energy/momentum.

The phenomenology of $R$-parity breaking models has been also 
studied.  If either $B$-violating or $L$-violating terms exist in 
Eq.~(\ref{eq:WMSSM}), but not both, they would not induce proton 
decay.\cite{HS}  However they can still produce $n$-$\bar{n}$ 
oscillation and a plethora of flavor-changing phenomena.  We refer to 
a recent compilation of phenomenological constraints\cite{Dreiner} 
for further details.

\subsection{Soft Supersymmetry Breaking Terms}

In addition to the interactions that arise from the superpotential 
Eq.~(\ref{eq:WMSSM}), we should add soft supersymmetry breaking terms 
to the Lagrangian as we have not seen any of the superpartners of the 
Standard Model particles.  Following the general classifications in 
Eq.~(\ref{eq:soft}), and assuming $R$-parity conservation, they are 
given by
\begin{eqnarray}
        & & {\cal L}_{\it soft} = {\cal L}_{1} + {\cal L}_{2}, \\
        & & {\cal L}_{1} = 
        - m_{Q}^{2ij} \tilde{Q}_{i}^{*} \tilde{Q}_{j}
        - m_{U}^{2ij} \tilde{U}_{i}^{*} \tilde{U}_{j}
        - m_{D}^{2ij} \tilde{D}_{i}^{*} \tilde{D}_{j} \nonumber \\
        & & \qquad - m_{L}^{2ij} \tilde{L}_{i}^{*} \tilde{L}_{j}
        - m_{E}^{2ij} \tilde{E}_{i}^{*} \tilde{E}_{j}
        - m_{H_{u}}^{2} |H_{u}|^{2} - m_{H_{d}}^{2} |H_{d}|^{2}, 
        \label{eq:softm2}\\
        & & {\cal L}_{2} = 
        - A_{u}^{ij} \lambda_{u}^{ij} \tilde{Q}_{i} \tilde{U}_{j} H_{u}
        - A_{d}^{ij} \lambda_{d}^{ij} \tilde{Q}_{i} \tilde{D}_{j} H_{d}
        - A_{l}^{ij} \lambda_{e}^{ij} \tilde{Q}_{i} \tilde{U}_{j} H_{d}
        + B \mu H_{u} H_{d} + c.c.
        \nonumber \\
        \label{eq:trilinear}
\end{eqnarray}
The mass-squared parameters for scalar quarks (squarks) and scalar 
leptons (sleptons) are all three-by-three hermitian matrices, while 
the trilinear couplings $A^{ij}$ and the bilinear coupling $B$ of mass 
dimension one are general complex numbers.\footnote{It is unfortunate 
that the notation $A$ is used both for the scalar components of chiral 
supermultiplets and the trilinear couplings.  Hopefully one can tell 
them apart from the context.}

\subsection{Higgs Sector}

It is of considerable interest to look closely at the Higgs sector of 
the MSSM.  Following the general form of the supersymmetric 
Lagrangians Eqs.~(\ref{eq:Kahler2},\ref{eq:VF},\ref{eq:VD}) with the 
superpotential $W = \mu H_{u} H_{d}$ in Eq.~(\ref{eq:WMSSM}) as well as 
the soft parameters in Eq.~(\ref{eq:softm2}), the potential for the 
Higgs bosons is given as
\begin{eqnarray}
        & & V = \frac{g^{\prime 2}}{2} \left( H_{u}^{\dagger} \frac{1}{2} H_{u}
        + H_{d}^{\dagger} \frac{-1}{2} H_{d} \right)^{2}
        + \frac{g^{2}}{2} \left( H_{u}^{\dagger} \frac{\vec{\tau}}{2} H_{u}
        + H_{d}^{\dagger} \frac{\vec{\tau}}{2} H_{d} \right)^{2}
        \nonumber \\
        & & 
        + \mu^{2} (|H_{u}|^{2} + |H_{d}|^{2})
        + m_{H_{u}}^{2} |H_{u}|^{2} + m_{H_{d}}^{2} |H_{d}|^{2}
        - (B \mu H_{u} H_{d} + c.c.)
        \label{eq:VMSSM}
\end{eqnarray}
It turns out that it is always possible to gauge-rotate the Higgs 
bosons such that
\begin{equation}
        \langle H_{u} \rangle 
        = \left( \begin{array}{c} 0 \\ v_{u} \end{array} \right),
        \qquad
        \langle H_{d} \rangle
        \left( \begin{array}{c} v_{d} \\ 0 \end{array} \right),
        \label{eq:HVEV}
\end{equation}
in the vacuum.  Since only electrically neutral components have 
vacuum expectation values, the vacuum necessarily conserves $U(1)_{\rm 
QED}$.\footnote{This is not necessarily true in general two-doublet 
Higgs Models.  Consult a review.\cite{hunters}}  Writing the 
potential (\ref{eq:VMSSM}) down using the expectation values 
(\ref{eq:HVEV}), we find
\begin{equation}
        V = \frac{g_{Z}^{2}}{8} (v_{u}^{2} - v_{d}^{2})^{2}
        + (v_{u}\ v_{d}) 
        \left( \begin{array}{cc}
                \mu^{2} + m_{H_{u}}^{2} & - B \mu\\
                - B \mu & \mu^{2} + m_{H_{d}}^{2}
        \end{array} \right)
        \left( \begin{array}{c}
                v_{u} \\ v_{d}
        \end{array} \right),
        \label{eq:VMSSM2}
\end{equation}
where $g_{Z}^{2} = g^{2} + g^{\prime 2}$.  In order for the Higgs 
bosons to acquire the vacuum expectation values, the determinant of 
the mass matrix at the origin must be negative,
\begin{equation}
        {\rm det} \left( \begin{array}{cc}
                \mu^{2} + m_{H_{u}}^{2} & - B \mu\\
                - B \mu & \mu^{2} + m_{H_{d}}^{2}
        \end{array} \right)
        < 0.
        \label{eq:constraint1}
\end{equation}
However, there is a danger that the direction $v_{u} = v_{d}$, which 
makes the quartic term in the potential identically vanish, may be 
unbounded from below.  For this not to occur, we need
\begin{equation}
        \mu^{2} + m_{H_{u}}^{2} + \mu^{2} + m_{H_{d}}^{2} > 2 \mu B.
        \label{eq:constraint2}
\end{equation}

In order to reproduce the mass of the $Z$-boson correctly, 
we need
\begin{equation}
        v_{u} = \frac{v}{\sqrt{2}} \sin \beta, \qquad
        v_{d} = \frac{v}{\sqrt{2}} \cos \beta, \qquad
        v = 250~{\rm GeV}.
        \label{eq:vac}
\end{equation}
The vacuum minimization conditions are given by $\partial V/\partial 
v_{u} = \partial V/\partial v_{d} = 0$ from the potential 
Eq.~(\ref{eq:VMSSM2}).  Using Eq.~(\ref{eq:vac}), we obtain
\begin{equation}
        \mu^{2} = - \frac{m_{Z}^{2}}{2}
        + \frac{m_{H_{d}}^{2} - m_{H_{u}}^{2} \tan^{2} \beta}{\tan^{2} \beta 
        -1},
        \label{eq:mu}
\end{equation}
and
\begin{equation}
        B \mu = (2\mu^{2} + m_{H_{u}}^{2} + m_{H_{d}}^{2}) \sin\beta \cos\beta.
        \label{eq:Bmu}
\end{equation}

Because there are two Higgs doublets, each of which with four real 
scalar fields, the number of degrees of freedom is eight before the 
symmetry breaking.  However three of them are eaten by $W^{+}$, 
$W^{-}$ and $Z$ bosons, and we are left with five physics scalar 
particles.  There are two CP-even scalars $h^{0}$, $H^{0}$, one CP-odd 
scalar $A^{0}$, and two charged scalars $H^{+}$ and $H^{-}$.  Their 
masses can be worked out from the potential (\ref{eq:VMSSM2}):
\begin{equation}
        m_{A}^{2} = 2\mu^{2} + m_{H_{u}}^{2} + m_{H_{d}}^{2},
        \qquad
        m_{H^{\pm}}^{2} = m_{W}^{2} + m_{A}^{2},
        \label{eq:Higgsmass1}
\end{equation}
and 
\begin{equation}
        m_{h^{0}}^{2}, m_{H^{0}}^{2} = \frac{1}{2}
        \left( m_{A}^{2} + m_{Z}^{2} 
        \pm \sqrt{(m_{A}^{2} + m_{Z}^{2})^{2} - 4 m_{Z}^{2} m_{A}^{2} \cos^{2} 
        2\beta} \right).
        \label{eq:Higgsmass2}
\end{equation}
A very interesting consequence of the formula 
Eq.~(\ref{eq:Higgsmass2}) is that the lighter CP-even Higgs mass 
$m_{h^{0}}^{2}$ is maximized when $\cos^{2} 2\beta = 1$: 
$m_{h^{0}}^{2} = (m_{A}^{2} + m_{Z}^{2} - |m_{A}^{2} - m_{Z}^{2}|)/2$.  
When $m_{A} < m_{Z}$, we obtain $m_{h^{0}}^{2} = m_{A}^{2} < 
m_{Z}^{2}$, while when $m_{A} > m_{Z}$, $m_{h^{0}}^{2} = m_{Z}^{2}$.  
Therefore in any case we find 
\begin{equation}
        m_{h^{0}} \leq m_{Z}.
        \label{eq:Higgsbound}
\end{equation}
This is an important prediction in the MSSM. The reason why the masses 
of the Higgs boson are related to the gauge boson masses is that the 
Higgs quartic couplings in Eq.~(\ref{eq:VMSSM}) are all determined by 
the gauge couplings because they originate from the elimination of the 
auxiliary $D$-fields in Eq.~(\ref{eq:Kahler2}).

Unfortunately, the prediction Eq.~(\ref{eq:Higgsbound}) is modified at 
the one-loop level,\cite{1-loop} approximately as
\begin{equation}
        \Delta (m_{h^{0}}^{2}) = \frac{N_{c}}{4\pi^{2}}
        h_{t}^{4} v^{2} \sin^{4}\beta \log 
        \left(\frac{m_{\tilde{t}_{1}}m_{\tilde{t}_{2}}}{m_{t}^{2}}\right).
        \label{eq:Deltamh0}
\end{equation}
With the scalar top mass of up to 1~TeV, the lightest Higgs mass is 
pushed up to about 130~GeV.  (See also the latest analysis including 
resummed two-loop contribution.\cite{2-loop})

\begin{figure}
        \centerline{ \psfig{file=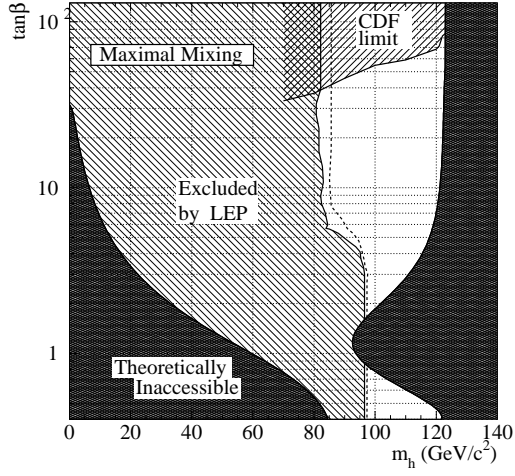,width=0.5\textwidth} } 
        \caption{Regions in the $(m_{h^{0}}, \tan\beta)$ plane excluded by 
        the MSSM Higgs boson searches at LEP in data up to 189~GeV, and at 
        CDF in run I data.  The regions not allowed by the MSSM for a top 
        mass of 175~GeV, a SUSY scale of 1~TeV and maximal mixing 
        in the stop sector are also indicated.  The dotted curve is the 
        LEP expected limit.\protect\cite{LP99Higgs}}
        \label{fig:MSSMHiggs}
\end{figure}

The parameter space of the MSSM Higgs sector can be described by two 
parameters.  This is because the potential Eq.~(\ref{eq:VMSSM2}) has 
three independent parameters, $\mu^{2} + m_{H_{u}}^{2}$, $\mu^{2} + 
m_{H_{d}}^{2}$, and $B \mu$, while one combination is fixed by the 
$Z$-mass Eq.~(\ref{eq:constraint1}).  It is customary to pick 
either $(m_{A}, \tan\beta)$, or $(m_{h^{0}}, \tan\beta)$ to present 
experimental constraints.  The current experimental constraint on 
this parameter space is shown in Fig.~\ref{fig:MSSMHiggs}.\footnote{The 
large $\tan\beta$ region may appear completely excluded in the plot, 
but this is somewhat misleading; it is due to the parameterization 
$(m_{h^{0}}, \tan\beta)$ which squeezes the $m_{h^{0}}$ region close 
to the theoretical upper bound to a very thin one.  In the $(m_{A}, 
\tan\beta)$ parameterization, one can see the allowed region much more 
clearer.}

\begin{figure}
        \centerline{ \psfig{file=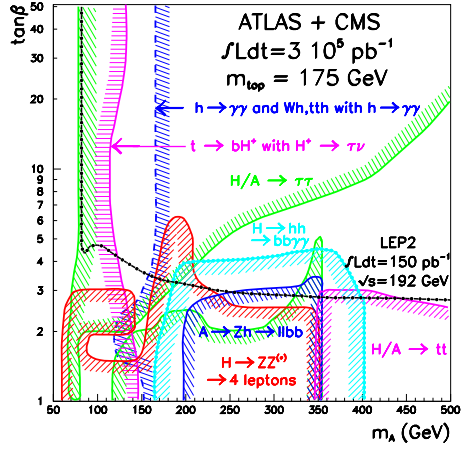,width=0.5\textwidth} } 
        \caption{Expected coverage of the MSSM Higgs sector parameter space 
        by the LHC experiments, after three years of high-luminosity 
        running.}
        \label{fig:LHC-MSSM}
\end{figure}

The range of the Higgs mass predicted in the MSSM is not necessarily 
an easy range for the LHC experiments, but three-years' running at the 
high luminosity is supposed to cover the entire MSSM parameter space, 
by employing many different production/decay modes as seen in 
Fig.~\ref{fig:LHC-MSSM}.

\subsection{Neutralinos and Charginos}

Once the electroweak symmetry is broken, and since supersymmetry is
already explicitly broken in the MSSM, there is no quantum number
which can distinguish two neutral higgsino states 
$\tilde{H}_{u}^{0}$, $\tilde{H}_{d}^{0}$, and two neutral gaugino 
states $\tilde{W}^{3}$ (neutral wino) and $\tilde{B}$ (bino).  They 
have four-by-four Majorana mass matrix
\begin{eqnarray}
    \lefteqn{
    {\cal L} \supset -\frac{1}{2}
    \times } \nonumber \\
    & & (\tilde{B}\ \tilde{W}^{3}\ \tilde{H}_{d}^{0}\ \tilde{H}_{u}^{0})
    \left( \begin{array}{cccc}
        M_{1} & 0 
        & - m_{Z}s_{W}c_{\beta} & m_{Z}s_{W}s_{\beta}\\
        0 & M_{2}
        & m_{Z}c_{W}c_{\beta} & - m_{Z}c_{W}s_{\beta}\\
        - m_{Z}s_{W}c_{\beta} & m_{Z}c_{W}c_{\beta}
        & 0 & -\mu\\
        m_{Z}s_{W}s_{\beta} & - m_{Z}c_{W}s_{\beta}
        & -\mu & 0
        \end{array} \right)
        \left( \begin{array}{c}
        \tilde{B}\\ \tilde{W}^{3}\\ \tilde{H}_{d}^{0}\\ \tilde{H}_{u}^{0}
        \end{array}\right).\nonumber \\
    \label{eq:neutralinos}
\end{eqnarray}
Here, $s_{W} = \sin \theta_{W}$, $c_{W} = \cos \theta_{W}$, $s_{\beta}
= \sin \beta$, and $c_{\beta} = \cos \beta$.  Once $M_{1}$, $M_{2}$,
$\mu$ exceed $m_{Z}$, which is preferred given the current
experimental limits, one can regard components proportional to $m_{Z}$
as small perturbations.  Then the neutralinos are close to their weak
eigenstates, bino, wino, and higgsinos.  But the higgsinos in this
limit are mixed to form symmetric and anti-symmetric linear
combinations $\tilde{H}_{S}^{0} =
(\tilde{H}_{d}^{0}+\tilde{H}_{u}^{0})/\sqrt{2}$ and $\tilde{H}_{A}^{0}
= (\tilde{H}_{d}^{0}-\tilde{H}_{u}^{0})/\sqrt{2}$.

Similarly two positively charged inos: $\tilde{H}_{u}^{+}$ and 
$\tilde{W}^{+}$, and two negatively charged inos: $\tilde{H}_{d}^{-}$ 
and $\tilde{W}^{-}$ mix.  The mass matrix is given by
\begin{equation}
    {\cal L}\supset
    - (\tilde{W}^{-}\ \tilde{H}_{d}^{-})
    \left( \begin{array}{cc}
    M_{2} & \sqrt{2} m_{W} s_{\beta} \\
    \sqrt{2} m_{W} c_{\beta} & \mu
    \end{array} \right)
    \left( \begin{array}{c}
    \tilde{W}^{+} \\ \tilde{H}_{u}^{+}
    \end{array} \right) + c.c.
    \label{eq:charginos}
\end{equation}
Again once $M_{2}, \mu \gtrsim m_{W}$, the chargino states are close 
to the weak eigenstates winos and higgsinos.  

\subsection{Squarks, Sleptons}

The mass terms of squarks and sleptons are also modified after the
electroweak symmetry breaking.  There are four different
contributions.  One is the supersymmetric piece coming from the
$|\partial W/\partial \phi_{i}|^{2}$ terms in Eq.~(\ref{eq:VF}) with
$\phi_{i} = Q, U, D, L, E$.  These terms add $m_{f}^{2}$ where $m_{f}$
is the mass of the quarks and leptons from their Yukawa couplings to
the Higgs boson.  Next one is combing from the $|\partial W/\partial
\phi_{i}|^{2}$ terms in Eq.~(\ref{eq:VF}) with $\phi_{i} = H_{u}$ or
$H_{d}$ in the superpotential Eq.~(\ref{eq:WMSSM}).  Because of the
$\mu$ term,
\begin{eqnarray}
    \frac{\partial W}{\partial H_{u}^{0}}
    & = & -\mu H_{d}^{0} + \lambda_{u}^{ij} \tilde{Q}_{i} \tilde{U}_{j},
    \label{eq:dWdHu}  \\
    \frac{\partial W}{\partial H_{d}^{0}} 
    & = & -\mu H_{d}^{0} + \lambda_{d}^{ij} \tilde{Q}_{i} \tilde{D}_{j}
    + \lambda_{e}^{ij} \tilde{L}_{i} \tilde{E}_{j}.
    \label{eq:dWdHd}
\end{eqnarray}
Taking the absolute square of these two expressions pick the cross 
terms together with $\langle H_{d}^{0} \rangle = v 
\cos\beta/\sqrt{2}$, $\langle H_{u}^{0} \rangle = v 
\sin\beta/\sqrt{2}$ and we obtain mixing between $\tilde{Q}$ and 
$\tilde{U}$, $\tilde{Q}$ and $\tilde{D}$, and $\tilde{L}$ and 
$\tilde{E}$.  Similarly, the vacuum expectation values of the Higgs 
bosons in the trilinear couplings Eq.~(\ref{eq:trilinear}) also 
generate similar mixing terms.  Finally, the $D$-term potential after 
eliminating the auxiliary field $D$ Eq.~(\ref{eq:VD}) also give 
contributions to the scalar masses $m_{Z}^{2} (I_{3} - Q \sin^{2} 
\theta_{W}) \cos 2\beta$.  Therefore, the mass matrix of stop, for 
instance, is given as
\begin{eqnarray}
    \lefteqn{
    {\cal L} \supset
    - (\tilde{t}_{L}^{*}\ \tilde{t}_{R}^{*})} \nonumber \\
    & & 
    \left( \begin{array}{cc}
    m_{Q_{3}}^{2} + m_{t}^{2} 
    + m_{Z}^{2} ( \frac{1}{2} - \frac{2}{3} s^{2}_{W}) c_{2\beta}
    & m_{t} (A_{t} - \mu \cot\beta) \\
    m_{t} (A_{t} - \mu \cot\beta) 
    & m_{U_{3}}^{2} + m_{t}^{2} 
    + m_{Z}^{2} ( - \frac{2}{3} s^{2}_{W}) c_{2\beta}
    \end{array} \right)
    \left( \begin{array}{c}
    \tilde{t}_{L} \\ \tilde{t}_{R}
    \end{array} \right), \nonumber \\
    \label{eq:stop}
\end{eqnarray}
with $c_{2\beta} = \cos 2\beta$.  Here, $\tilde{t}_{L}$ is the up 
component of $\tilde{Q}_{3}$, and $\tilde{t}_{R} = \tilde{T}^{*}$.  
For first and second generation particles, the off-diagonal terms are 
negligible for most purposes.  They may, however, be important when 
their loops in flavor-changing processes are considered.

\subsection{What We Gained in the MSSM}

It is useful to review here what we have gained in the MSSM over what 
we had in the Standard Model.  The main advantage of the MSSM is of 
course what motivated the supersymmetry to begin with: the absence of 
the quadratic divergences as seen in Eq.~(\ref{eq:deltamH}).  This 
fact allows us to apply the MSSM down to distance scales much shorter 
than the electroweak scale, and hence we can at least hope that many 
of the puzzles discussed at the beginning of the lecture to be solved 
by physics at the short distance scales.  

There are a few amusing and welcome by-products of supersymmetry 
beyond this very motivation.  First of all, the Higgs doublet in the 
Standard Model appears so unnatural partly because it is the only 
scalar field introduced just for the sake of the electroweak symmetry 
breaking.  In the MSSM, however, there are so many scalar fields: 15 
complex scalar fields for each generation and two in each Higgs 
doublet.  Therefore, the Higgs bosons are just ``one of them.''  Then 
the question about the electroweak symmetry breaking is addressed in a 
completely different fashion: why is it only the Higgs bosons that
condense?  In fact, one can even partially answer this question in the 
renormalization group analysis in the next sections where 
``typically'' (we will explain what we mean by this) it is only the 
Higgs bosons which acquire negative mass squared 
(\ref{eq:constraint1}) while the masses-squared of all the other 
scalars ``naturally'' remain positive.  Finally, the absolute upper 
bound on the lightest CP-even Higgs boson is falsifiable by 
experiments.  

However, life is not as good as we wish.  We will see that there 
are very stringent low-energy constraints on the MSSM in the next 
section.

\section{Low-Energy Constraints}

Despite the fact that we are interested in superparticles in the 
100--1000~GeV range, which we are just starting to explore in collider 
searches, there are many amazingly stringent low-energy constraints on 
superparticles.

One of the most stringent constraints comes from the 
$K^{0}$--$\bar{K}^{0}$ mixing parameters $\Delta m_{K}$ and 
$\varepsilon_{K}$.  The main reason for the stringent constraints is that 
the scalar masses-squared in the MSSM Lagrangian Eq.~(\ref{eq:softm2}) 
can violate flavor, {\it i.e.}\/, the scalar masses-squared matrices 
are not necessarily diagonal in the basis where the corresponding 
quark mass matrices are diagonal.

\begin{figure}
        \centerline{ \psfig{file=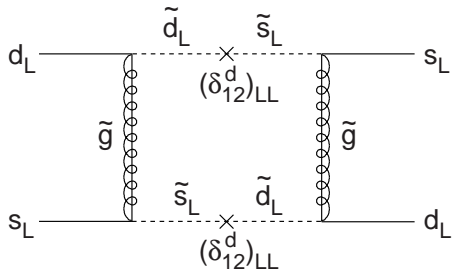,width=0.5\textwidth} } 
        \caption{A Feynman diagram which gives rise to $\Delta m_{K}$ and 
        $\varepsilon_{K}$.}
        \label{fig:gluinobox}
\end{figure}

To simplify the discussion, let us concentrate only on the first and the 
second generations (ignore the third).  We also go to the basis where 
the down-type Yukawa matrix $\lambda_{d}^{ij}$ is diagonal, such that
\begin{equation}
        \lambda_{d}^{ij} v_{d} = \left( \begin{array}{cc}
                m_{d} & 0 \\ 0 & m_{s}
        \end{array} \right).
        \label{eq:md}
\end{equation}
Therefore the states $K^{0} = (d\bar{s})$, $\bar{K}^{0} = (s\bar{d})$ 
are well-defined in this basis.  In the same basis, however, the 
squark masses-squared can have off-diagonal elements in general,
\begin{equation}
        m_{Q}^{2ij} = \left( \begin{array}{cc}
                m_{\tilde{d}_{L}}^{2} & m_{Q, 12}^{2}\\
                m_{Q, 12}^{2*} & m_{\tilde{s}_{L}}^{2}
        \end{array}
        \right),
        \qquad
        m_{D}^{2ij} = \left( \begin{array}{cc}
                m_{\tilde{d}_{R}}^{2} & m_{D, 12}^{2}\\
                m_{D, 12}^{2*} & m_{\tilde{s}_{R}}^{2}
        \end{array}
        \right).
        \label{eq:mdtilde}
\end{equation}
Since their off-diagonal elements will be required to be small (as we 
will see later), it is convenient to treat them as small perturbation. 
We insert the off-diagonal elements as two-point Feynman vertices 
which change the squark flavor $\tilde{d}_{L,R}\leftrightarrow 
\tilde{s}_{L,R}$ in the diagrams.  To simplify the discussion further, 
we assume that all squarks and gluino the are comparable in their 
masses $\tilde{m}$.  Then the relevant quantities are given in terms 
of the ratio $(\delta^{d}_{12})_{LL} \equiv m_{Q, 
12}^{2}/\tilde{m}^{2}$ (and similarly $(\delta^{d}_{12})_{RR} = m_{D, 
12}^{2}/\tilde{m}^{2}$), as depicted in Fig.~\ref{fig:gluinobox}.  The 
operator from this Feynman diagram is estimated approximately as
\begin{equation}
        0.005 \alpha_{s}^{2} \frac{(\delta_{12}^{d})_{LL}^{2}}{\tilde{m}^{2}}
        (\bar{d}_{L} \gamma^{\mu} s_{L})(\bar{d}_{L} \gamma_{\mu} s_{L}).
        \label{eq:KKoperator}
\end{equation}
This operator is further sandwiched between $K^{0}$ and $\bar{K}^{0}$ 
states, and we find
\begin{eqnarray}
        & &\Delta m_{K}^{2} \sim 0.005 f_{K}^{2} m_{K}^{2} \alpha_{s}^{2} 
        (\delta_{12}^{d})_{LL}^{2} \frac{1}{\tilde{m}^{2}}
        \nonumber \\
        & & = 1.2 \times 10^{-12}~{\rm GeV}^{2}
        \left( \frac{f_{K}}{160~{\rm MeV}}\right)^{2}
        \left( \frac{\alpha_{s}}{0.1} \right)^{2}       
        (\delta_{12}^{d})_{LL}^{2}
        < 3.5 \times 10^{-15}~{\rm GeV}^{2}, \nonumber \\
        \label{eq:DmKestimate}
\end{eqnarray}
where the last inequality is the phenomenological constraint in the 
absence of accidental cancellations.  This requires
\begin{equation}
        (\delta_{12}^{d})_{LL} \lesssim 0.05 
        \left( \frac{\tilde{m}}{500~{\rm GeV}} \right)
        \label{eq:deltalimit}
\end{equation}
and hence the off-diagonal element $m_{Q, 12}^{2}$ must be small.  It 
turns out that the product $(\delta_{12}^{d})_{LL} 
(\delta_{12}^{d})_{RR}$ is more stringently constrained, especially its 
imaginary part from $\varepsilon_{K}$.  Much more careful and detailed 
analysis than the above order-of-magnitude estimate 
gives\cite{Ciuchini}
\begin{equation}
        {\rm Re} \left[(\delta_{12}^{d})_{LL} 
        (\delta_{12}^{d})_{RR}\right] < (0.016)^{2},
        \qquad
        {\rm Im} \left[(\delta_{12}^{d})_{LL} 
        (\delta_{12}^{d})_{RR}\right] < (0.0022)^{2}.
        \label{eq:Ciuchini}
\end{equation}

There are many other low-energy observables, such as electron and 
neutron electric dipole moments (EDM), $\mu \rightarrow e \gamma$, 
which place important constraints on the supersymmetry 
parameters.\cite{Masieroreview}

There are various ways to avoid such low-energy constraints on 
supersymmetry.  The first one is called ``universality'' of soft 
parameters.\cite{DG} It is simply assumed that the scalar 
masses-squared matrices are proportional to identity matrices, {\it 
i.e.}\/, $m_{Q}^{2}, m_{U}^{2}, m_{D}^{2} \propto {\bf 
1}$.  Then no matter what rotation is made in order to go to the basis 
where the quark masses are diagonal, the identity matrices stay the 
same, and hence the off-diagonal elements are never produced.  There 
has been many proposals to generate universal scalar masses either by 
the mediation mechanism of the supersymmetry breaking such as the 
gauge mediated (see reviews\cite{GR}), anomaly 
mediated\cite{anomalymediation}, or gaugino 
mediated\cite{gauginomediation} supersymmetry breaking, or by 
non-Abelian flavor symmetries.\cite{nonabelian} The second possibility 
is called ``alignment,'' where certain flavor symmetries should be 
responsible for ``aligning'' the quark and squark mass matrices such
that the squark masses are almost diagonal in the same basis where 
the down-quark masses are diagonal.\cite{alignment} Because of the CKM 
matrix it is impossible to do this both for down-quark and up-quark 
masses.  Since the phenomenological constraints in the up-quark sector 
are much weaker than in the down-quark sector, this choice would 
alleviate many of the low-energy constraints (except for 
flavor-diagonal CP-violation such as EDMs).  Finally there is a 
possibility called ``decoupling,'' which assumes first- and 
second-generation superpartners much heavier than TeV while keeping 
the third-generation superpartners as well as gauginos in the 100~GeV 
range to keep the Higgs self-energy small enough.\cite{decoupling}  
Even though this idea suffers from a fine-tuning problem in 
general,\cite{nondecoupling} many models had been constructed to 
achieve such a split mass spectrum recently.\cite{moreminimal}

In short, the low-energy constraints are indeed very stringent, but 
there are many ideas to avoid such constraints naturally within 
certain model frameworks.  Especially given the fact that we still do 
not know any of the superparticle masses experimentally, one cannot 
make the discussions more clear-cut at this stage.  On the other hand, 
important low-energy effects of supersymmetry are still being 
discovered in the literature, such as muon $g-2$,\cite{g-2} and direct 
CP-violation.\cite{epsilonp} They may be even more possible low-energy 
manifestations of supersymmetry which have been missed so far.

\section{Renormalization Group Analyses}

Once supersymmetry protects the Higgs self-energy against corrections 
from the short distance scales, or equivalently, the high energy 
cutoff scales, it becomes important to connect physics at the 
electroweak scale where we can do measurements to the fundamental 
parameters defined at high energy scales.  This can be done by 
studying the renormalization-group evolution of parameters.  It also 
becomes a natural expectation that the supersymmetry breaking itself 
originates at some high energy scale.  If this is the case, the soft 
supersymmetry breaking parameters should also be studied using the 
renormalization-group equations.  We study the renormalization-group 
evolution of various parameters in the softly-broken supersymmetric 
Lagrangian at the one-loop level.\footnote{Recently, there have been 
developments in obtaining and understanding all-order beta 
functions for gauge coupling constants\cite{NSVZ} and soft 
parameters.\cite{allorder}}  If supersymmetry indeed turns out to 
be the choice of nature, the renormalization-group analysis will be 
crucial in probing physics at high energy scales using the 
observables at the TeV-scale collider experiments.\cite{probing}

\subsection{Gauge Coupling Constants}

The first parameters to be studied are naturally the coupling 
constants in the Standard Model.  The running of the gauge couplings 
constants are described in term of the beta functions, and their 
one-loop solutions in non-supersymmetric theories are given by
\begin{equation}
        \frac{1}{g^{2}(\mu)} = \frac{1}{g^{2}(\mu')} 
        + \frac{b_{0}}{8\pi^{2}} \log \frac{\mu}{\mu'},
        \label{eq:gaugeRGE}
\end{equation}
with
\begin{equation}
        b_{0} = \frac{11}{3} C_{2} (G) - \frac{2}{3} S_{f} - \frac{1}{3} S_{b}.
        \label{eq:gaugebeta}
\end{equation}
This formula is for Weyl fermions $f$ and complex scalars $b$.  The 
group theory factors are defined by
\begin{eqnarray}
        \delta^{ad} C_{2}(G) & = & f^{abc} f^{dbc}
        \label{eq:C2}  \\
        \delta^{ab} S_{f,b} & = & {\rm Tr} T^{a} T^{b}
        \label{eq:Sf}
\end{eqnarray}
and $C_{2}(G) = N_{c}$ for SU($N_{c}$) groups and $S_{f,b}=1/2$ for 
their fundamental representations.  

In supersymmetric theories, there is always the gaugino multiplet in 
the adjoint representation of the gauge group.  They contribute to 
Eq.~(\ref{eq:gaugebeta}) with $S_{f} = C_{2}(G)$, and therefore the total 
contribution of the vector supermultiplet is $3 C_{2} (G)$.  On the 
other hand, the chiral supermultiplets have a Weyl spinor and a 
complex scalar, and the last two terms in Eq.~(\ref{eq:gaugebeta}) are 
always added together to $S_{f}=S_{b}$.  Therefore, the beta function 
coefficients simplify to
\begin{equation}
        b_{0} = 3 C_{2} (G) - S_{f} .
        \label{eq:gaugebetaSUSY}
\end{equation}
Given the beta functions, it is easy to work out how the gauge 
coupling constants measured accurately at LEP/SLC evolve to higher 
energies.

\begin{figure}
        \centerline{\psfig{file=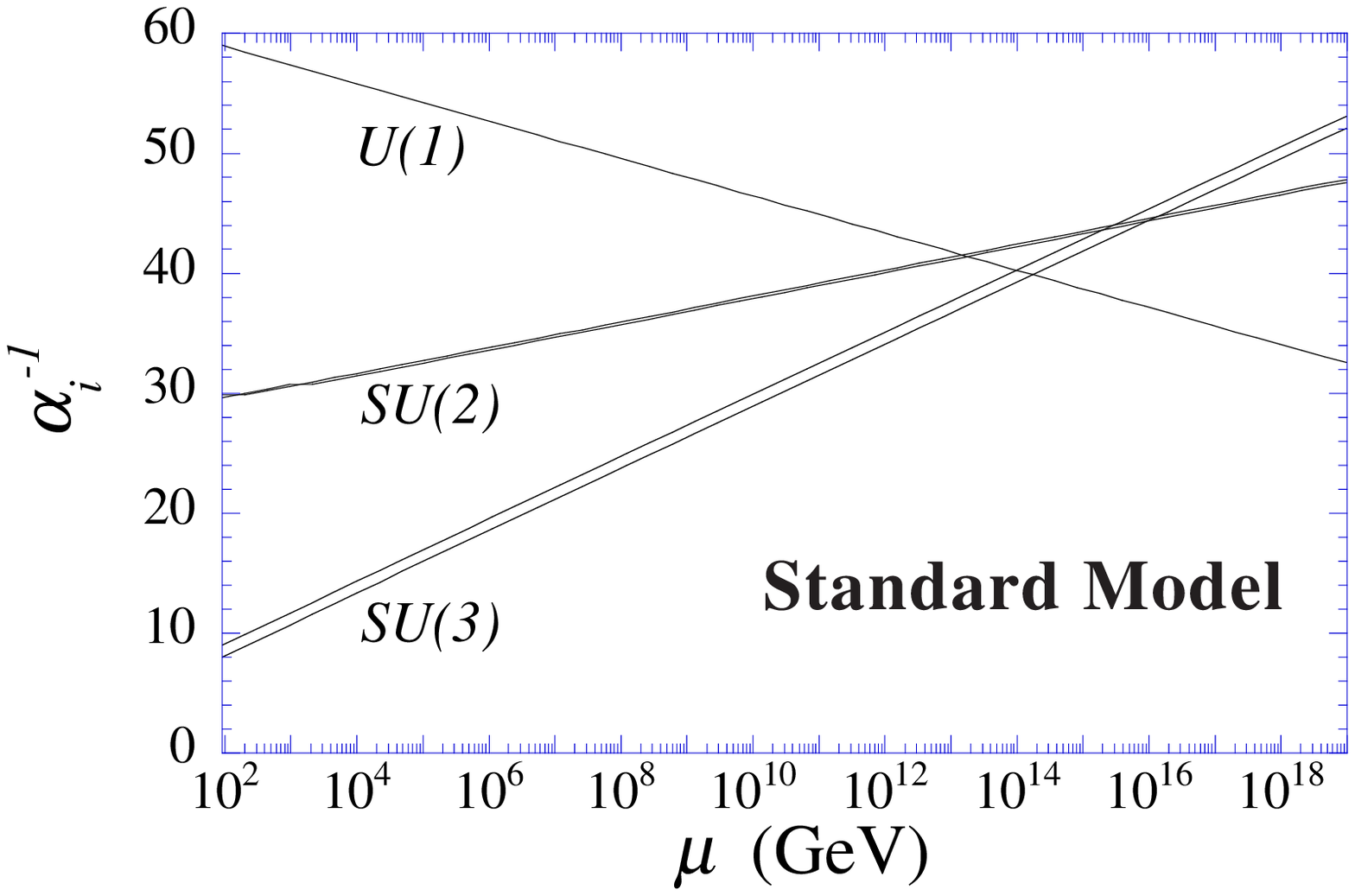,width=0.5\textwidth}
        \psfig{file=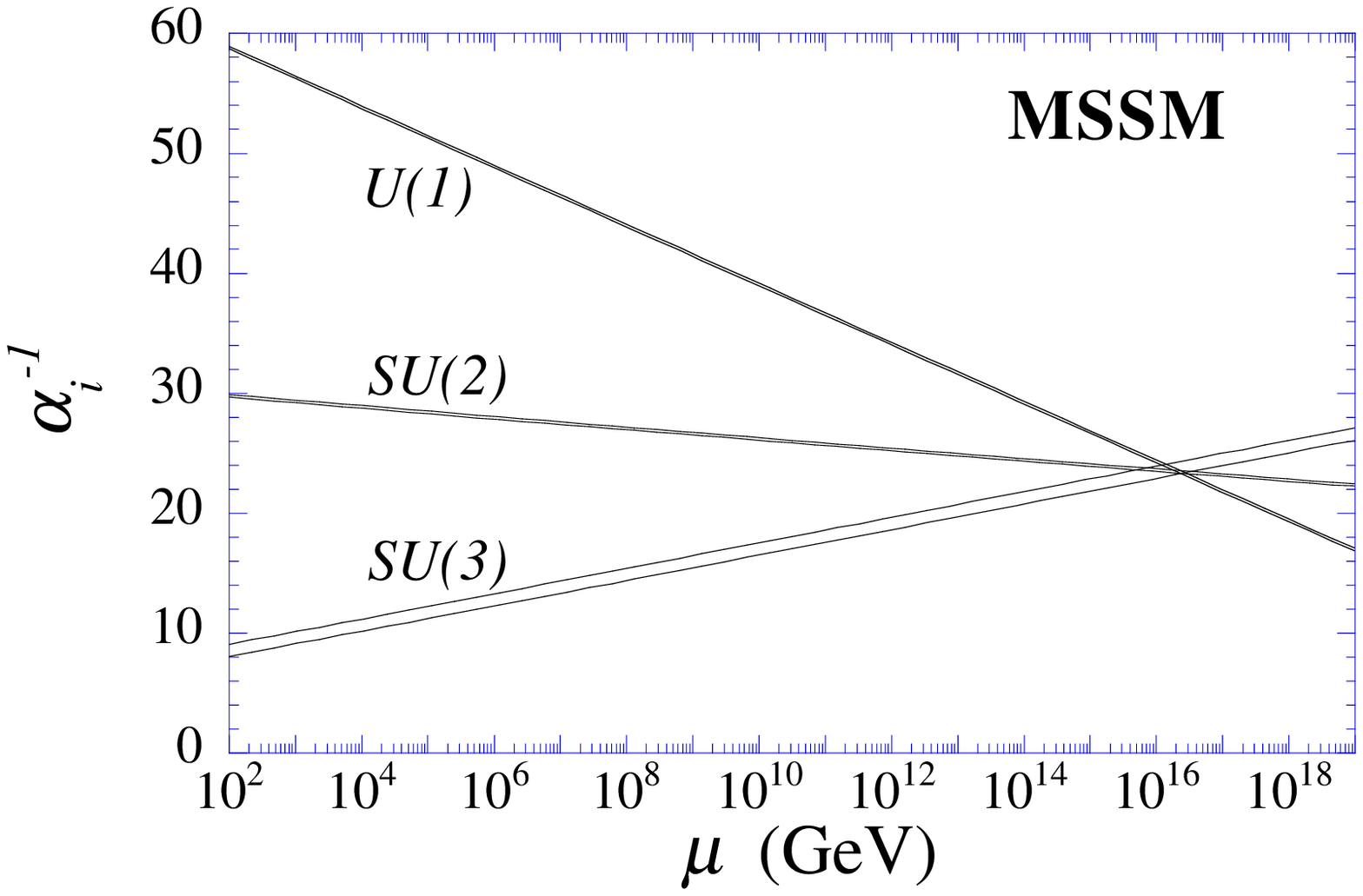,width=0.5\textwidth}}
        \caption{Running of gauge coupling constants in the Standard Model 
        and in the MSSM.}
        \label{fig:GUT}
\end{figure}

One interesting possibility is that the gauge groups in the Standard 
Model $SU(3)_{C} \times SU(2)_{L} \times U(1)_{Y}$ may be embedded 
into a simple group, such as $SU(5)$ or $SO(10)$, at some high energy 
scale, called ``grand unification.''  The gauge coupling constants at 
$\mu \sim m_{Z}$ are approximately $\alpha^{-1} = 129$, $\sin^{2} 
\theta_{W} \simeq 0.232$, and $\alpha_{s}^{-1} = 0.119$.  In the 
$SU(5)$ normalization, the $U(1)$ coupling constant is given by 
$\alpha_{1} = \frac{5}{3} \alpha' = \frac{5}{3} \alpha/\cos^{2} 
\theta_{W}$.  It turns out that the gauge coupling constants become 
equal at $\mu \simeq 2\times 10^{16}$~GeV given the MSSM particle 
content (Fig.~\ref{fig:GUT}).  On the other hand, the three gauge 
coupling constants miss each other quite badly with the 
non-supersymmetric Standard Model particle content.  This observation 
suggests the possibility of supersymmetric grand unification.

\subsection{Yukawa Coupling Constants}

\begin{figure}
        \centerline{\psfig{file=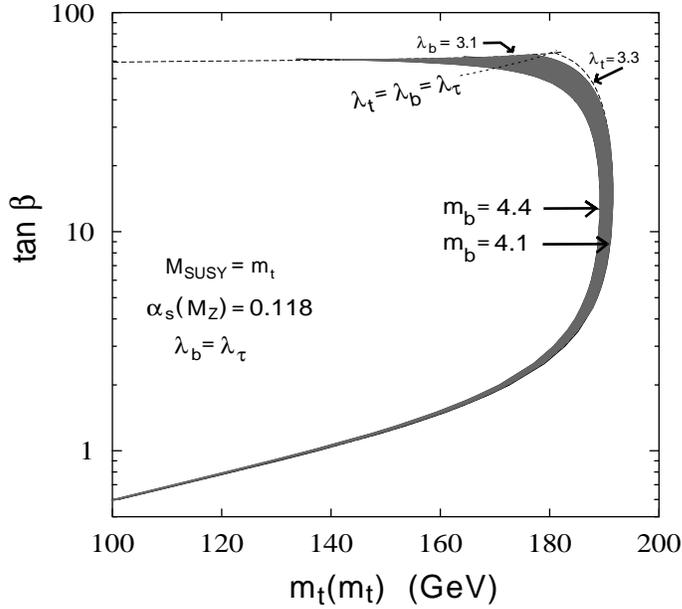,width=0.8\textwidth}}
        \caption{The regions on $(m_{t}, \tan\beta)$ plane where 
        $h_{b}=h_{\tau}$ at the GUT-scale.\protect\cite{BBOP}}
        \label{fig:Yukawa}
\end{figure}

Since first- and second-generation Yukawa couplings are so small, let 
us ignore them and concentrate on the third-generation ones.  Their 
renormalization-group equations are given as
\begin{eqnarray}
        \mu \frac{d h_{t}}{d\mu} & = & \frac{h_{t}}{16\pi^{2}}
        \left[6 h_{t}^{2} + h_{b}^{2} - \frac{16}{3} g_{3}^{2}
        - 3 g_{2}^{2} - \frac{13}{15} g_{1}^{2} \right],
        \label{eq:runht}  \\
        \mu \frac{d h_{b}}{d\mu} & = & \frac{h_{b}}{16\pi^{2}}
        \left[6 h_{b}^{2} + h_{t}^{2} + h_{\tau}^{2} - \frac{16}{3} g_{3}^{2}
        - 3 g_{2}^{2} - \frac{7}{15} g_{1}^{2} \right],
        \label{eq:runhb}  \\
        \mu \frac{d h_{\tau}}{d\mu} & = & \frac{h_{\tau}}{16\pi^{2}}
        \left[4 h_{\tau}^{2} + 3 h_{b}^{2} 
        - 3 g_{2}^{2} - \frac{9}{5} g_{1}^{2} \right].
        \label{eq:runhtau}
\end{eqnarray}
The important aspect of these equations is that the gauge coupling 
constants push down the Yukawa coupling constants at higher energies, 
while the Yukawa couplings push them up.  This interplay, together with 
a large top Yukawa coupling, allows the possibility that the Yukawa 
couplings may also unify at the same energy scale where the gauge 
coupling constants appear to unify (Fig.~\ref{fig:Yukawa}).  It turned 
out that the actual situation is much more relaxed than what this plot 
suggests.  This is because there is a significant correction to 
$m_{b}$ at $\tan \beta \gtrsim 10$ when the superparticles are integrated 
out \cite{HRS}.

\subsection{Soft Parameters}

Since we do not know any of the soft parameters at this point, we 
cannot use the renormalization-group equations to probe physics at 
high energy scales.  On the other hand, we can use the 
renormalization-group equations from boundary conditions at high 
energy scales suggested by models to obtain useful information on the 
``typical'' superparticle mass spectrum.  

First of all, the gaugino mass parameters have very simple behavior 
that
\begin{equation}
        \mu \frac{d}{d \mu} \frac{M_{i}}{g_{i}^{2}} = 0.
        \label{eq:gauginoRGE}
\end{equation}
Therefore, the ratios $M_{i}/g_{i}^{2}$ are constants at all 
energies.  If the grand unification is true, both the gauge coupling 
constants and the gaugino mass parameters must unify at the GUT-scale 
and hence the ratios are all the same at the GUT-scale.  Since the 
ratios do not run, the ratios are all the same at any energy scales, 
and hence the low-energy gaugino mass ratios are predicted to be
\begin{equation}
        M_{1} : M_{2} : M_{3} = g_{1}^{2} : g_{2}^{2} : g_{3}^{2}
        \sim 1 : 2 : 7
        \label{eq:gauginoratio}
\end{equation}
at the TeV scale.  We see the tendency that the colored particle 
(gluino in this case) is much heavier than uncolored particle (wino 
and bino in this case).  This turns out to be a relatively 
model-independent conclusion.

The running of scalar masses is given by simple equations when all 
Yukawa couplings other than that of the top quark are neglected.  We 
find
\begin{eqnarray}
        16 \pi^{2} \mu \frac{d}{d\mu} m^{2}_{H_{u}} & = & 
        3 X_{t} - 6 g_{2}^{2} M_{2}^{2} - \frac{6}{5} g_{1}^{2} M_{1}^{2},
        \label{eq:runmHu}  \\
        16 \pi^{2} \mu \frac{d}{d\mu} m^{2}_{H_{d}} & = & 
        - 6 g_{2}^{2} M_{2}^{2} - \frac{6}{5} g_{1}^{2} M_{1}^{2},
        \label{eq:runmHd}  \\
        16 \pi^{2} \mu \frac{d}{d\mu} m^{2}_{Q_{3}} & = & 
        X_{t} - \frac{32}{3} g_{3}^{2} M_{3}^{2} - 6 g_{2}^{2} M_{2}^{2} 
        - \frac{2}{15} g_{1}^{2} M_{1}^{2},
        \label{eq:runmQ3}  \\
        16 \pi^{2} \mu \frac{d}{d\mu} m^{2}_{U_{3}}  & = & 
        2 X_{t} - \frac{32}{3} g_{3}^{2} M_{3}^{2} 
        - \frac{32}{15} g_{1}^{2} M_{1}^{2}.
        \label{eq:runmU3}
\end{eqnarray}
Here, $X_{t} = 2 h_{t}^{2} (m_{H_{u}}^{2} + m_{Q_{3}}^{2} + 
m_{U_{3}}^{2})$ and the trilinear couplings are also neglected.  Even 
within this simplified assumptions, one learns interesting lessons.  
First of all, the gauge interactions push the scalar masses up at 
lower energies due to the gaugino mass squared contributions.  Colored 
particles are pushed up even more than uncolored ones, and the 
right-handed sleptons would be the least pushed up.  On the other hand, 
Yukawa couplings push the scalar masses down at lower energies.  The 
coefficients of $X_{t}$ in the Eqs.~(\ref{eq:runmHu}, \ref{eq:runmQ3}, 
\ref{eq:runmU3}) are simply the multiplicity factors which correspond 
to 3 of $SU(3)_{C}$, 2 of $SU(2)_{Y}$ and 1 of $U(1)_{Y}$.  It is 
extremely amusing that the $m^{2}_{H_{u}}$ is pushed down the most 
because of the factor of three as well as is pushed up the least 
because of the absence of the gluino mass contribution.  Therefore, 
the fact that the Higgs mass squared is negative at the electroweak 
scale may well be just a simple consequence of the 
renormalization-group equations!  Since the Higgs boson is just ``one 
of them'' in the MSSM, the renormalization-group equations provide a 
very compelling reason why it is only the Higgs boson whose 
mass-squared goes negative and condenses.  One can view this as an 
explanation for the electroweak symmetry breaking.

\subsection{Minimal Supergravity}

Of course, nothing quantitative can be said unless one makes some 
specific assumptions for the boundary conditions of the 
renormalization-group equations.  One common choice called ``Minimal 
Supergravity'' is the following set of assumptions:
\begin{eqnarray*}
        & & m_{Q}^{2ij} = m_{U}^{2ij} = m_{D}^{2ij} = m_{L}^{2ij}
        = m_{E}^{2ij} = m_{0}^{2} \delta^{ij},
         \\
        & & m_{H_{u}}^{2} = m_{H_{d}}^{2} = m_{0}^{2},  \\
        & & A_{u}^{ij} = A_{d}^{ij} = A_{l}^{ij} = A_{0}
         \\
        & & M_{1} = M_{2} = M_{3} = M_{1/2} 
\end{eqnarray*}
at the GUT-scale.  The parameter $m_{0}$ is called the universal 
scalar mass, $A_{0}$ the universal trilinear coupling, and $M_{1/2}$ 
the universal gaugino mass.  Once this assumption is made, there are 
only five parameters at the GUT-scale, $(m_{0}, M_{1/2}, A_{0}, B, 
\mu)$.  This assumption also avoids most of the low-energy constraints 
easily because the scalar mass-squared matrices are proportional to 
the identity matrices and hence there is no flavor violation.  Of 
course this is probably an oversimplification of the parameter space, 
but it still provides useful starting point in discussing 
phenomenology.  Especially most of the search limits from collider 
experiments have been reported using this assumption.  In general, 
this choice of the boundary conditions, which actually have not much 
to do with supergravity itself, lead to acceptable and interesting 
phenomenology including the collider signatures, low-energy 
constraints as well as cosmology.

\section{Collider Phenomenology}

We do not go into much details of the collider phenomenology of 
supersymmetry in this lecture notes and we refer to reviews.\cite{DPF} 
Here, we give only a very brief summary of collider phenomenology.  
Supersymmetry is an ideal target for current and new future collider 
searches.  As long as they are within the mass scale expected by the 
argument given at the beginning of the lecture, we expect 
supersymmetric particles to be discovered at LEP-II (even though the 
phase space left is quite limited by now), Tevatron Run-II, or the 
LHC.

\begin{figure}
\centerline{\psfig{file=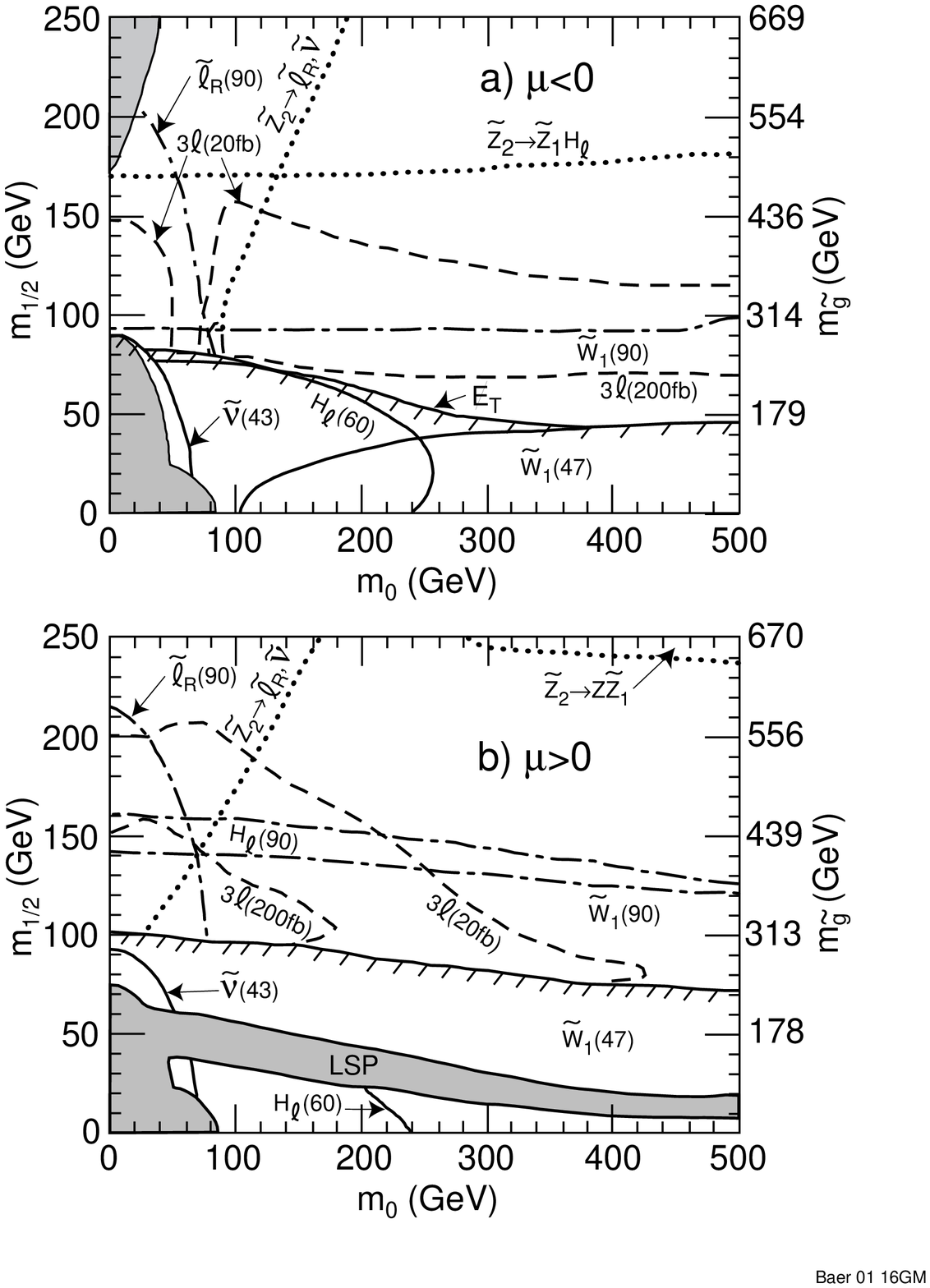,height=11cm}}
\caption[]{Regions in the $m_{0}\ vs. m_{1/2}$ plane explorable by
Tevatron and LEP II experiments.\protect\cite{DPF}}
\label{fig:LEP-Tevatron}
\end{figure}

\begin{figure}
\centerline{\psfig{file=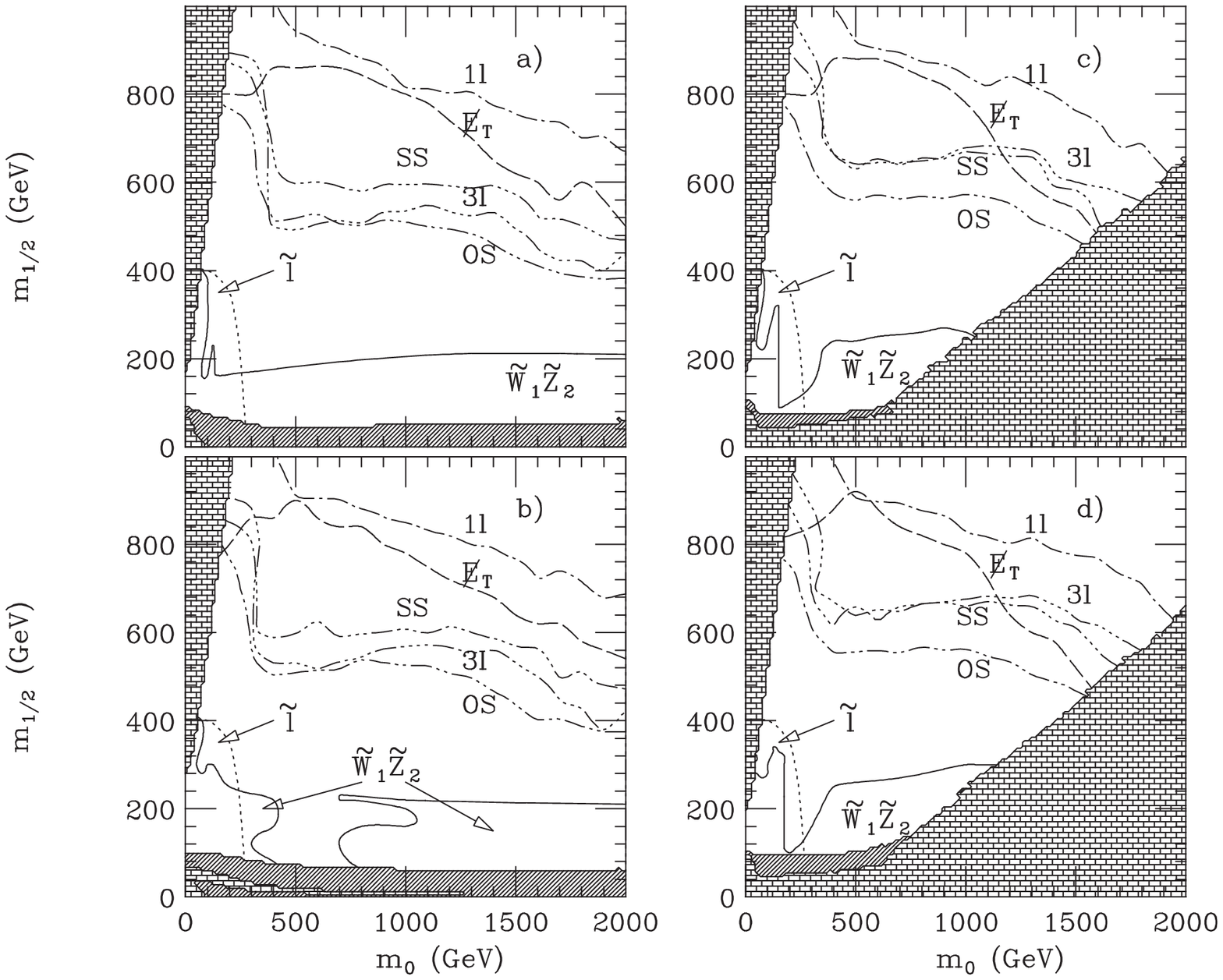,height=11cm}}
\caption[]{Regions in the $m_{0}\ vs. m_{1/2}$ plane explorable by
LHC experiments with 10~fb$^{-1}$ of integrated 
luminosity.\protect\cite{LHCreach}  Different curves correspond to 
different search modes: 1l (single lepton), $\not\!\!E_{T}$ (missing 
transverse energy), SS (same sign dilepton), 3l (trilepton), OS 
(opposite sign dilepton), $\tilde{l}$ (slepton), $\tilde{W}_{1} 
\tilde{Z}_{2}$ (charged wino, neutral wino associated production).}
\label{fig:LHC}
\end{figure}

The next two figures Figs.~\ref{fig:LEP-Tevatron}, \ref{fig:LHC} show
the discovery reach of supersymmetry at LEP-II, Tevatron Run II, LHC.
It is fair to say that the mass range of superparticles relevant to
solve the problem of fine cancellation in the Higgs boson self-energy
described at the beginning of the lecture is covered by these 
experiments.  

\begin{figure}
    \centerline{
    \psfig{file=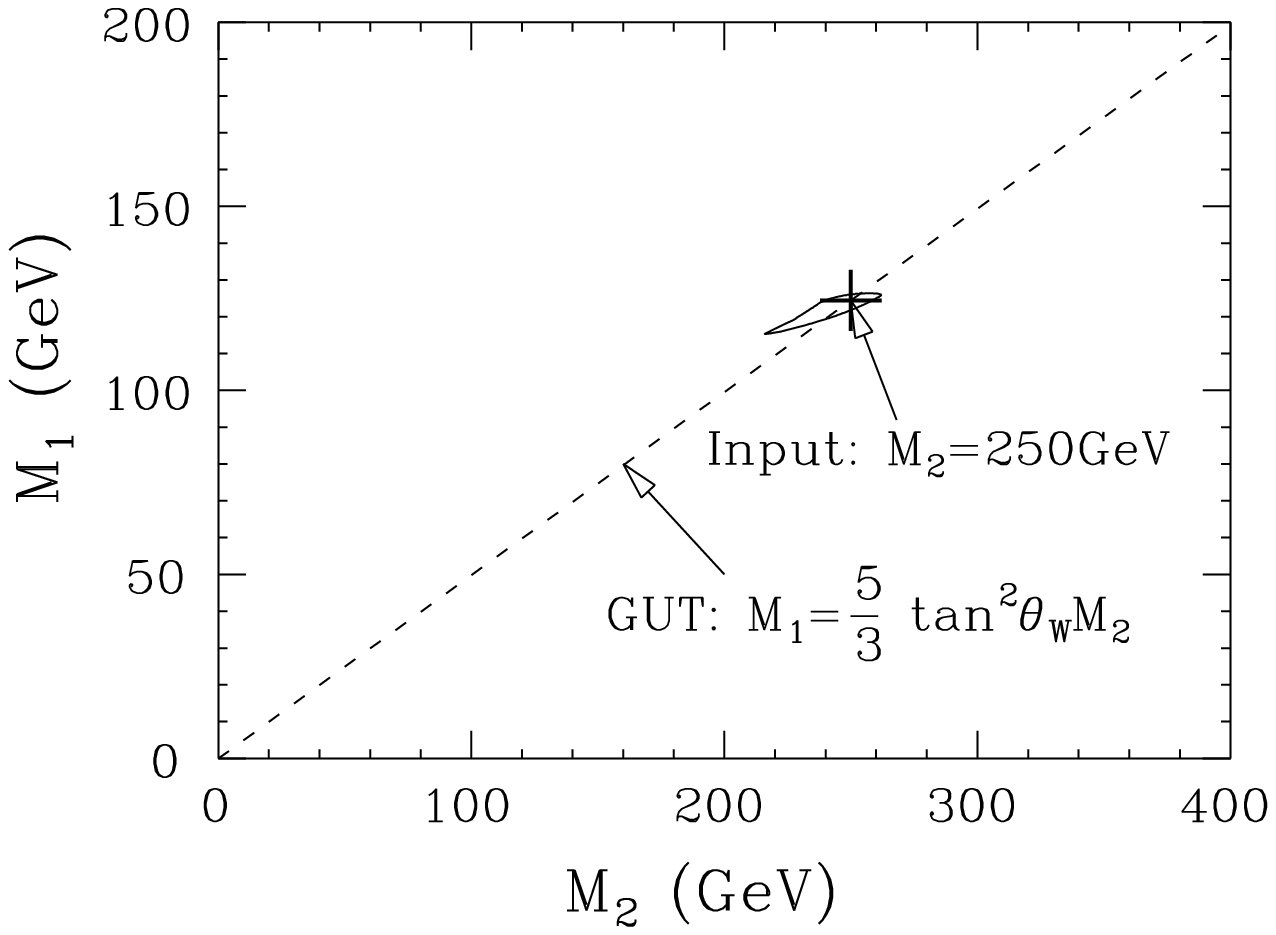,width=0.6\textwidth}
    }
    \centerline{
     \psfig{file=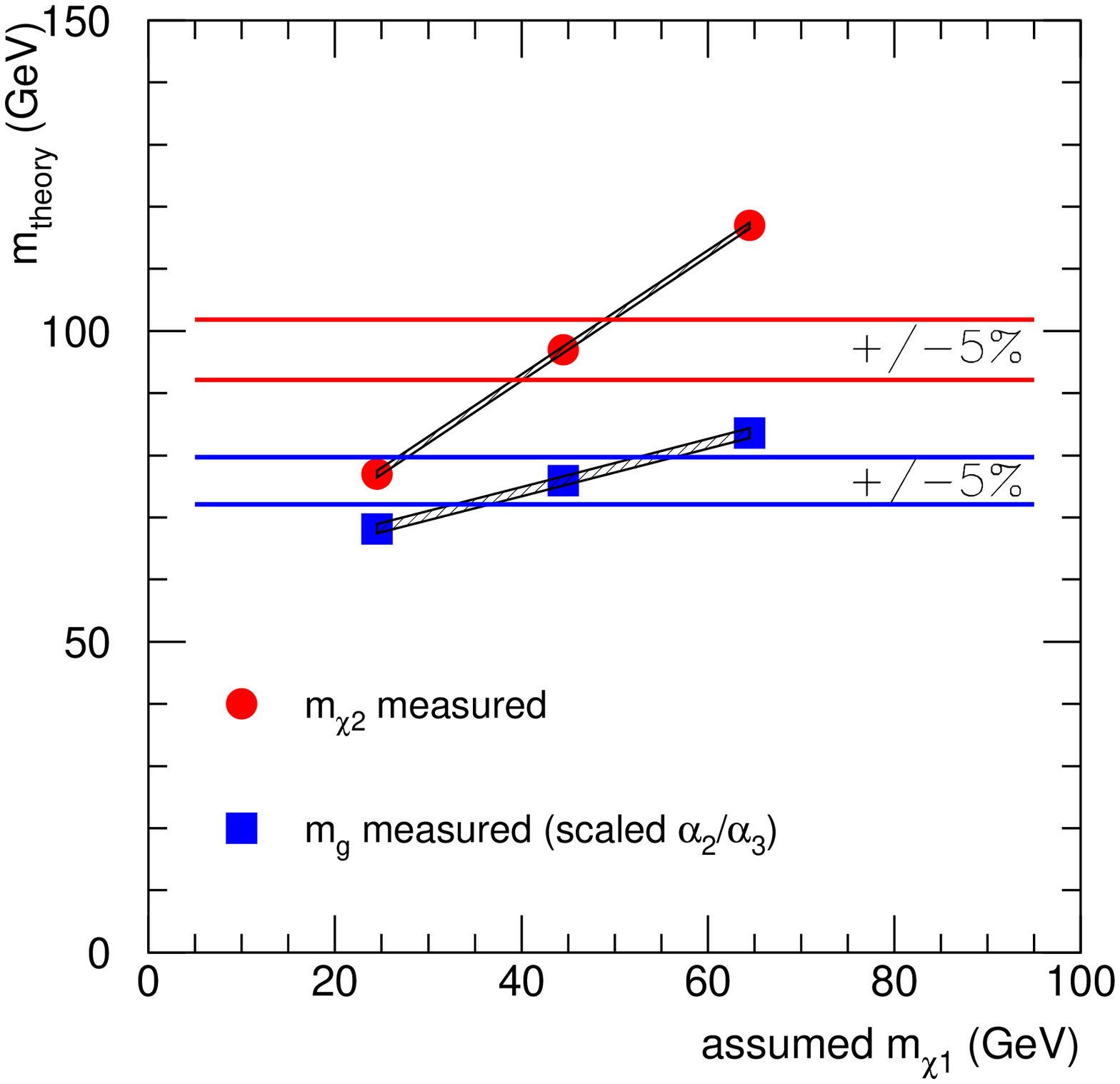,width=0.6\textwidth}
     }
     \caption{Experimental tests of gaugino mass unification at a
    future $e^+ e^-$ collider\,\protect\cite{Tsukamoto} and the
    LHC.\,\protect\cite{Marjie}}
    \label{fig:GUTtest}
\end{figure}

A future $e^{+} e^{-}$ linear collider would play a fantastic role in
proving that new particles are indeed superpartners of the known
Standard Model particles and in determining their parameters.\cite{Tsukamoto}  
Once such studies will be done, we will exploit
renormalization-group analyses trying to connect physics at TeV scale
to yet-more-fundamental physics at higher energy scales.  Example of 
such possible studies are shown in Fig.~\ref{fig:GUTtest}.  The 
measurements of gaugino masses were simulated.  At the LHC, the 
measurements are basically on the gluino mass and the LSP mass which is 
assumed to be the bino state, and their mass difference can be 
measured quite well.  By assuming a value of the LSP mass, one can 
extract the gluino mass.  At the $e^{+} e^{-}$ linear colliders, one 
can even disentangle the mixing in neutralino and chargino states 
employing expected high beam polarizations and determine $M_{1}$ and 
$M_{2}$ in a model-independent matter.  Combination of both types of 
experiments determine all three gaugino masses, which would provide a 
non-trivial test of the grand unification.

\begin{figure}
\centerline{\psfig{file=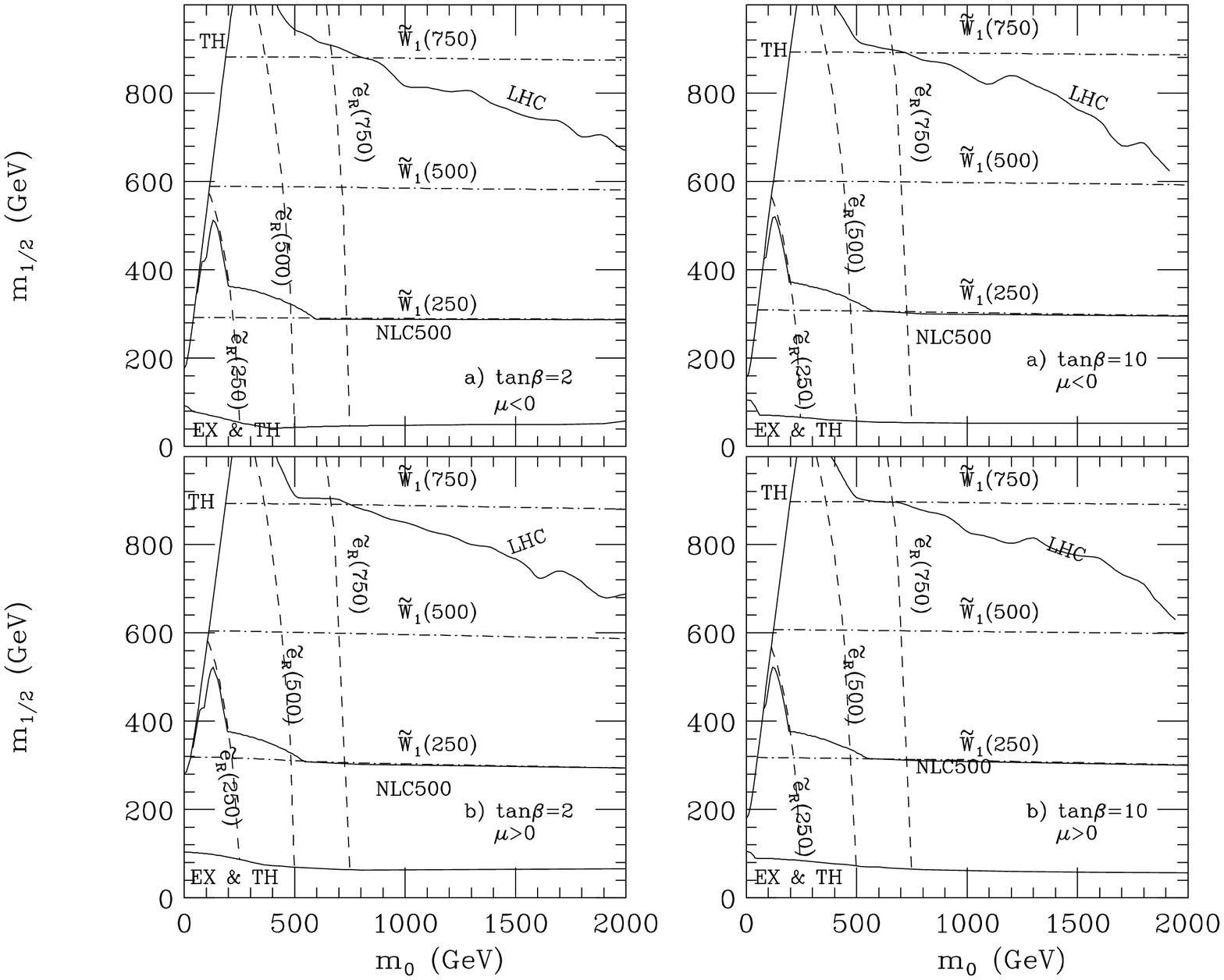,height=11cm}}
\caption[]{Regions in the $m_{0}\ vs. m_{1/2}$ plane explorable by
$e^{+} e^{-}$ linear collider experiments with 20~fb$^{-1}$ of 
integrated luminosity.}
\label{fig:NLC}
\end{figure}

\section{Mediation Mechanisms of Supersymmetry Breaking}

One of the most important questions in the supersymmetry 
phenomenology is how supersymmetry is broken and how the particles in 
the MSSM learn the effect of supersymmetry breaking.  The first one is 
the issue of dynamical supersymmetry breaking, and the second one is 
the issue of the ``mediation'' mechanism.  

The problem of the supersymmetry breaking itself has gone through a
dramatic progress in the last few years thanks to works on the
dynamics of supersymmetric gauge theories by Seiberg.\cite{IS} The
original idea by Witten\cite{motivation} was that the dynamical
supersymmetry breaking is ideal to explain the hierarchy.  Because of
the non-renormalization theorem, if supersymmetry is unbroken at the
tree-level, it remains unbroken at all orders in perturbation theory. 
However, they may be non-perturbative effects suppressed by
$e^{-8\pi^{2}/g^{2}}$ that could break supersymmetry.  Then the energy
scale of the supersymmetry breaking can be naturally suppressed
exponentially compared to the energy scale of the fundamental theory
(string?).  Even though this idea attracted a lot of
interest,\footnote{I didn't live through this era, so this is just a
guess.} the model building was hindered by the lack of understanding in
dynamics of supersymmetric gauge theories.  Only relative few models
were convincingly shown to break supersymmetry dynamically, such as
the $SU(5)$ model with two pairs\cite{SU(5)x2} of ${\bf 5}^{*} + {\bf
10}$ and the 3-2 model.\cite{3-2} After Seiberg's works,
however, there has been an explosion in the number of models which
break supersymmetry dynamically (see a review\cite{AENSUSY97} and
references therein).  For instance, some of the models which were
claimed to break supersymmetry dynamically,
such as $SU(5)$ with one pair\cite{SU(5)x1} of ${\bf 5}^{*} + {\bf
10}$ or $SO(10)$ with one spinor\cite{SO(10)} ${\bf 16}$, are 
actually strongly coupled and could not be analyzed reliably (called
``non-calculable''), but new techniques allowed us to analyze these
strongly coupled models reliably.\cite{noncalculable} Unexpected
vector-like models were also found\cite{vectorlike} which proved to be
useful for model building.

There has also been an explosion in the number of mediation mechanisms 
proposed in the literature.  The oldest mechanism is that in 
supergravity theories where interactions suppressed by the Planck 
scale are responsible for communicating the effects of supersymmetry 
breaking to the particles in the MSSM. For instance, see a 
review.\cite{Nilles} Even though the gravity itself may not be the 
only effect for the mediation but there could be many operators 
suppressed by the Planck-scale responsible for the mediation, this 
mechanism was sometimes called ``gravity-mediation.''  The good thing 
about this mechanism is that this is almost always there.  However we 
basically do not have any control over the Planck-scale physics and 
the resulting scalar masses-squared are in general highly 
non-universal.  In this situation, the best idea is probably to 
constrain the scalar masses-squared matrix proportional to the 
identity matrix by non-Abelian flavor symmetries.\cite{nonabelian} 
Models were constructed where the breaking patterns of the flavor 
symmetry naturally explain the hierarchical quark and lepton mass 
matrices, while protecting the squark masses-squared matrices from 
deviating too far from the identity matrices.

A beautiful idea to guarantee the universal scalar masses is to use the
MSSM gauge interactions for the mediation.  Then the
supersymmetry breaking effects are mediated to the particles in the
MSSM in such a way that they do not distinguish particles in different
generations (``flavor-blind'') because they only depend on the gauge
quantum numbers of the particles.  Such a model was regarded difficult
to construct in the past.\cite{3-2} However, a break-through was made
by Dine, Nelson and collaborators,\cite{gaugemediation} who started
constructing models where the MSSM gauge interactions could
indeed mediate the supersymmetry breaking effects, inducing postive
scalar masses-squared and large enough gaugino masses (which used to
be one of the most difficult things to achieve).\cite{gaugino} The
original models had three independent sectors, one for supersymmetry
breaking, one (the messenger sector) for mediation alone, and the last
one the MSSM.  Later models eliminated the messenger sector 
entirely\cite{direct} (see also reviews\cite{GMreviews}).

Difficulty still remained how large enough gaugino masses can be
generated in models where the sector of dynamical supersymmetry
breaking couples to the MSSM fields only by Planck-scale suppressed
interactions.\cite{gaugino} One could go around this problem by a
clever choice of the quantum numbers for a gauge singlet
field.\cite{singlet}  But it was not realized until recently that the
gaugino masses are generated by superconformal
anomaly.\cite{anomalymediation}  This observation was confirmed and 
further generalized by other groups.\cite{moreanomaly} 
Randall and Sundrum further realized
that one could even have scalar masses entirely from the
superconformal anomaly if the sector of dynamical supersymmetry
breaking and the MSSM particles are physically separated in the extra
dimensions.  The consequence was striking: the soft parameters were
determined solely by the low-energy theory and did not depend on the
physics at high energy scales at all.  This makes it attractive as a
solution to the problem of flavor-changing neutral currents, as the
low-energy interactions of first and second generations are indeed
nearly flavor-blind.  Even though such models initially suffered from
the problem that some of the scalars had negative mass-squared, simple
fixes were proposed.\cite{fixes} One can preserve the virtue of the
anomaly mediation, namely ultraviolet insensitivity, and construct realistic 
models.

Finally a new idea called ``gaugino mediation'' came out
lately.\cite{gauginomediation} This idea employs an extra dimension
where the gauge fields propagate in the bulk.  Supersymmetry is broken
on a different brane and the MSSM fields learn the supersymmetry
breaking effects by the MSSM gauge interactions, and hence solving the
flavor-changing problem.

\section{Conclusion}

Supersymmetry is a well-motivated candidate for physics beyond the 
Standard Model.  It would allow us to extrapolate the (supersymmetric 
version of the) Standard Model down to much shorter distances, giving 
us hope to connect the observables at TeV-scale experiments to 
parameters of the much more fundamental theories.  Even though it has 
been extensively studied over two decades, many new aspects of 
supersymmetry have been uncovered in the last few years.  We expect 
that research along this direction will continue to be fruitful.  We, 
however, really need a clear-cut confirmation (or falsification) 
experimentally.  The good news is that we expect it to be discovered, 
if nature did choose this direction, at the currently planned 
experiments.

\end{document}